\documentclass[
twocolumn,aps, pre,longbibliography,superscriptaddress
]{revtex4-2}
\usepackage{psfrag,epsfig,amsfonts,amssymb,amsmath,wasysym,bm}
\usepackage{dcolumn}
\usepackage{bbold}
\usepackage{xcolor}
\usepackage{tabularx, tabulary, booktabs}
\usepackage{longtable}
\usepackage{tabu}
\usepackage{mathtools}
\usepackage{bm}
\usepackage[normalem]{ulem}

\usepackage{enumitem} 

\usepackage{hyperref}

		\newcommand{\<}{\langle}	\renewcommand{\>}{\rangle}

\newcommand{\e}{\mathrm{e}}				\newcommand{\kB}{k_{\mathrm B}} 

\renewcommand{\d}{\mathrm{d}} \newcommand{\T}{\mathsf{T}}     \newcommand{\lmat}{\left( \begin{matrix}}	\newcommand{\rmat}{\end{matrix} \right)}

\newcommand{\mref}[1]{\ref{#1}}
\newcommand{\meqref}[1]{\eqref{#1}}
\newcommand{\sref}[1]{\ref{#1}}

\newcommand{\ta}{\tau_{\mathrm{a}}}
\newcommand{\Da}{D_{\mathrm{a}}}
\newcommand{\He}{\operatorname{He}}
\newcommand{\Pe}{\mathrm{Pe}}

\newcommand{\initial}{\mathrm{i}}
\newcommand{\final}{\mathrm{f}}

\definecolor{REcolor}{RGB}{0, 150, 0}

\begin{document}

\title{Ideal gas of active particles in a box}

\author{Siran Li}
\affiliation{School of Mathematical Sciences, Queen Mary University of London, London E1~4NS, UK}

\author{Ralf Eichhorn}
\affiliation{Nordita, Royal Institute of Technology and Stockholm University,
106 91 Stockholm, Sweden}

\author{Lennart Dabelow}
\affiliation{School of Mathematical Sciences, Queen Mary University of London, London E1~4NS, UK}

\date{\today}

\begin{abstract}
Extending a fundamental concept of thermodynamics to the nonequilibrium realm of active matter, we derive an ideal gas law
for a suspension
of noninteracting
active particles confined to a finite rectangular box. The law is exact for run-and-tumble particles (RTPs) in one dimension and 
established by
a systematic analytical expansion of the steady-state probability density for active Ornstein-Uhlenbeck particles in any dimension. It also agrees well with numerical results for active Brownian particles and RTPs in higher dimensions and recovers previously known relations
for the limit of large system sizes. Together, these results demonstrate the
applicability of the ideal active gas law to the three most widely studied theoretical models of colloidal active matter.
As an application, we analyze
thermodynamic processes involving active matter on a macroscopic level, without recourse to microscopic particle trajectories or self-propulsion fluctuations.
\end{abstract}

\maketitle

\paragraph*{Introduction.}
The ideal gas is a pivotal concept in thermodynamics and statistical mechanics \cite{LandauLifshitz:StatisticalPhysics, KittelKroemer:ThermalPhysics}.
It describes a system of $N$ identical, noninteracting point particles
that freely roam an accessible volume $V$ constrained by hard walls.
The particles' kinetic energy gives rise to the temperature $T$,
and collisions with the enclosing walls result in the pressure $P$.

Despite its simplicity, the ideal gas model captures essential properties of many real gases,
particularly at low density and high temperature.
Thanks to its simplicity, many of these properties can be described analytically.
A notable example is the ideal gas law,
an equation of state that establishes a connection between the system's thermodynamic state variables: $PV = N \kB T$,
where $\kB$ is Boltzmann's constant.

Recent years have seen great interest in generalizing the concepts of traditional thermodynamics to systems and processes away from equilibrium.
A particularly popular class of such systems is active matter \cite{Fodor:2022ibe, Romanczuk2012abp, Bechinger2016apc, Takatori2016fst},
whose defining property is that the constituting particles can take up energy from their environment and use it to move around autonomously and persistently.
Examples include bacteria, Janus colloids, and nanorobots.
At the micro- to nanometer scales
we are interested in here,
these particles typically live in a fluid environment at temperature $T$
and thus experience frictional and fluctuating forces from collisions with the surrounding molecules,
which result in diffusive motion characterized by the (thermal) diffusivity $D := \kB T / \gamma$ with $\gamma$ the Stokes friction coefficient.
Activity leads to a second source of erratic motion,
described by the active diffusivity $\Da$,
with the crucial difference that the corresponding fluctuations are persistent over a macroscopically relevant timescale $\ta$,
and do not obey a fluctuation-dissipation relation.
The relative strength of active vs.\ thermal fluctuations is quantified by the P\'eclet number $\Pe := \sqrt{\Da / D}$.

In this work, we develop the theory of an ideal gas of active particles in a box.
A central result is the ``ideal active gas law''
\begin{equation}
\label{eq:IdealActiveGasLaw}
    P V = N \kB T \, \frac{1 + \Pe^2}{1 + \Pe^2 \vartheta} \,,
\end{equation}
where $\vartheta := \frac{\tanh \ell}{\ell}$ and $\ell := L \sqrt{\frac{1 + \Pe^2}{4 D \ta}}$ is a dimensionless characteristic length scale of the system.
This result is exact for run-and-tumble particles (RTPs) in $d = 1$ dimension
and is the first-order approximation of a systematic expansion for active Ornstein-Uhlenbeck particles (AOUPs) in a hypercubic box $V = L^d$ in any dimension.
Numerically, we observe that it also reproduces well the behavior of active Brownian particles (ABPs) and RTPs in higher dimensions
$d>1$.
Without activity ($\Pe = 0$)
or in the limits of large persistence time ($\ta \to \infty$) or small box size ($L \to 0$, at fixed $N/V$), Eq.~\eqref{eq:IdealActiveGasLaw} reduces to the original ideal gas law for passive particles.
In the large-volume limit ($L \to \infty$ at fixed $N/V$),
it reduces to the relation $P V = N \kB T (1 + \Pe^2)$,
which is known to hold for AOUPs, RTPs, and ABPs alike \cite{Solon2015pin, Solon2015ppe, Winkler2015vps, MariniBettoloMarconi2015tsm, Cameron2023esa, Dabelow2025tni}.
This large-system-size result is commonly rationalized
by comparison with the ideal gas law, $PV=N\kB T_{\mathrm{eff}}$,
where $T_{\mathrm{eff}}=T(1+\Pe^2)$ is interpreted as an effective 
temperature that combines thermal and active fluctuations \cite{Solon2015pin, MariniBettoloMarconi2015tsm}.
However, as a key insight, our ideal active gas law
for finite system sizes, Eq.~\eqref{eq:IdealActiveGasLaw}
(see also Eqs.~\eqref{eq:Pk} and~\eqref{eq:IdealActiveGasLaw:multi}),
goes beyond this
interpretation:
It reveals
a non-trivial interplay between particle 
activity and the shape and dimensions of the confining box,
which is captured by the characteristic length scale $\ell$.
As a crucial benefit of our result, we can use \eqref{eq:IdealActiveGasLaw} (and \eqref{eq:Pk})
to describe thermodynamic processes at finite volume,
referring only to macroscopic state variables despite
the inherent non-equilibrium character of active matter.
In dimensions $d>1$, in particular, our results predict a box-shape dependence of thermodynamic quantities under volume changes,
which is absent in passive systems
[see Fig.~\ref{fig:Process}c].

In the following,
we derive the ideal active gas law~\eqref{eq:IdealActiveGasLaw} and various generalizations,
verify them against numerics for paradigmatic active particle models of AOUPs, RTPs, and ABPs,
and demonstrate how the law can be used to describe thermodynamic processes with active matter.

\paragraph*{Model.}
Our starting point is the overdamped Langevin equation for the position $\bm x = (x_1 \, \cdots \, x_d)^\T$ of a single active particle in $d$ dimensions,
\begin{equation}
\label{eq:EOM:x}
	\dot{\bm{x}}(t)
		= \bm v(t) + \sqrt{2D} \, \bm\xi(t) \,,
\end{equation}
confined to a rectangular box of volume $V = L_1 \cdots L_d$ with hard walls (no-flux boundary conditions).
The unbiased Gaussian white-noise processes $\bm\xi(t)$ with $\< \xi_i(t) \xi_j(t') \> = \delta_{ij} \delta(t - t')$
model the thermal fluctuations,
while $\bm v(t)$ is the fluctuating self-propulsion velocity.
For ABPs and RTPs, $\bm v(t) = v_0 \bm n(t)$,
where $v_0$ is the constant self-propulsion speed and $\bm n(t)$ is a unit vector with fluctuating orientation.
The latter follows rotational diffusion for ABPs,
whereas it is
sampled from a uniform distribution
on the $d$-dimensional unit sphere
at discrete, exponentially distributed tumbling times for RTPs;
the rate is $\ta^{-1}$ in both cases and $\Da = \< \bm v^2 \> \ta / d = v_0^2 \ta / d$.
For AOUPs, we write $\bm v(t) = \sqrt{2\Da} \,\bm\eta(t)$,
where $\bm\eta(t)$ is an Ornstein-Uhlenbeck process \cite{Risken1984fpe, Gardiner1985hsm} with correlation time $\ta$,
i.e., $\< \eta_i(t) \, \eta_j(t') \> = \delta_{ij} \e^{-\lvert t - t' \rvert / \ta} / 2\ta$.

\paragraph*{Steady-state solution.}
We first focus on the AOUP model and observe that the $d$ spatial coordinates are independent in this case (in contrast to RTPs and ABPs),
so we can restrict ourselves to $d = 1$
(using $x_1=x$, $\eta_1=\eta$).
We also temporarily switch to dimensionless variables $\tilde t = t / \ta$, $\tilde x = x / \sqrt{D \ta}$, and $\tilde\eta = \sqrt{2\ta} \eta$,
and immediately drop the tildes again until further notice.
The steady-state probability density $\rho(x, \eta)$ of the joint process of positions and active fluctuations satisfies the Fokker-Planck equation
\begin{equation}
\label{eq:FP}
	\frac{\partial^2 \rho}{\partial x^2} - \Pe \,\eta \frac{\partial\rho}{\partial x} + \frac{\partial^2 \rho}{\partial\eta^2} + \eta \frac{\partial\rho}{\partial\eta} + \rho = 0
\end{equation}
with reflecting boundary conditions in $x$ to implement the box's hard walls,
\begin{equation}
\label{eq:BC}
	\left. \frac{\partial\rho(x, \eta)}{\partial x} - \Pe \, \eta \, \rho(x, \eta) \right|_{x = 0, L} = 0 \,,
\end{equation}
and natural boundary conditions in $\eta$.
The only free parameters are the (dimensionless) box size $L$ and the Peclet number $\Pe$.

We solve Eqs.~\eqref{eq:FP} and~\eqref{eq:BC} by expanding $\rho$ into Hermite functions in $\eta$ \cite{Risken1984fpe, Caprini2018apc, GarciaMillan2021rtm},
\begin{equation}
\label{eq:rho:ansatz}
	\rho(x, \eta)
		= \frac{\e^{-\eta^2 / 2}}{\sqrt{2 \pi}} \sum_{n = 0}^\infty X_n(x) \He_n(\eta) \,,
\end{equation}
where $\He_n(\eta)$ is the $n$th probabilist's Hermite polynomial \cite{Abramowitz1972hmf},
\begin{equation}
	\He_n(\eta) = (-1)^n \e^{\eta^2 / 2} \frac{\d^n}{\d\eta^n} \e^{-\eta^2/2} \,.
\end{equation}
This can be interpreted as an expansion in the 
correlations that develop between $x$ and $\eta$ as a consequence of confinement.
It
leads to a linear system of ordinary differential equations (ODEs) for the $X_n(x)$,
\begin{equation}
\label{eq:ODE:X}
    X''_n - \Pe \left[ X'_{n-1} + (n+1) X'_{n+1} \right] - n \, X_n = 0 \,,
\end{equation}
with boundary conditions
\begin{equation}
\label{eq:BC:x}
    X'_n(x) - \Pe \left[ X_{n-1}(x) + (n+1) X_{n+1}(x) \right] = 0
\end{equation}
at $x = 0, L$.
Solutions of arbitrary accuracy can be constructed by truncating this system at some order $M$ and setting $X_{n} \equiv 0$ for all $n > M$.
The details of this construction are provided in Section~\sref{app:HermiteFunctionExpansion} of \cite{suppl_arxiv}.

\begin{figure}
\centering
\includegraphics[width=\columnwidth]{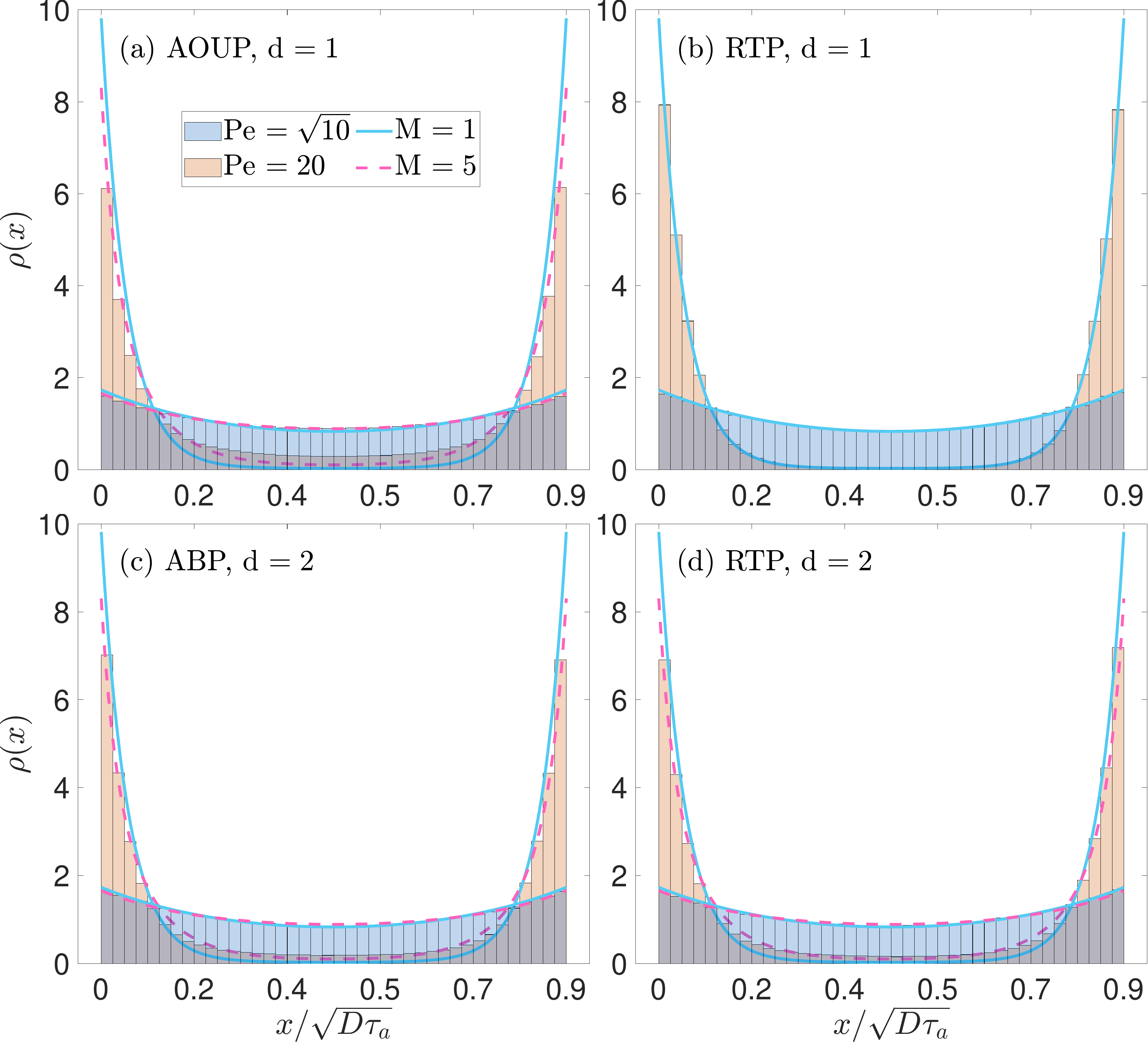}
\caption{
Marginal position distributions of various active particle models in a box of size $L / \sqrt{D \ta} = 0.9$ for different P\'eclet numbers $\Pe$ as indicated in the legend.
Histograms are obtained from numerical simulations using $5 \times 10^5$ steady-state position samples from $10^4$ independent trajectories.
Lines are theory curves $\rho(x) = X_0(x)$ [cf.\ Eq.~\eqref{eq:rho:x}] for different truncation orders $M$ (see legend).
(a) AOUP in $d = 1$.
(b) RTP in $d = 1$, the theory curve for $M=1$ [Eq.~\eqref{eq:rho:x:M1}] is exact.
(c) ABP in $d = 2$.
(d) RTP in $d = 2$.
In (c) and (d), the marginal distributions of a single coordinate ($x_1$) are shown.
}
\label{fig:PositionDistribution}
\end{figure}

As a first check, we inspect the marginal position distribution
\begin{equation}
\label{eq:rho:x}
    \rho(x) := \int_{-\infty}^\infty \d\eta \, \rho(x, \eta)
        = X_0(x) \,.
\end{equation}
In the last equality, we exploited the Hermite polynomials orthogonality property
\begin{equation}
\label{eq:He:Orthogonality}
    \int_{-\infty}^\infty \d\eta \, \frac{\e^{-\eta^2/2}}{\sqrt{2\pi}} \, \He_m(\eta) \, \He_n(\eta) = n! \, \delta_{mn}
\end{equation}
and the fact that $\He_0(\eta) = 1$.
Truncating at $M = 1$,
the marginal distribution is approximated by
\begin{equation}
\label{eq:rho:x:M1}
    \rho(x) = \frac{1 + \frac{\Pe^2}{\cosh\ell} \, \cosh\!\left(\sqrt{1+\Pe^2} [x - \frac{L}{2}]\right)}{L (1 + \Pe^2 \vartheta)}
\end{equation}
where $\vartheta$ and $\ell$ were defined below Eq.~\eqref{eq:IdealActiveGasLaw}.
In Fig.~\ref{fig:PositionDistribution},
we compare the
marginal position distributions $\rho(x)$,
Eq.~\eqref{eq:rho:x}, at different truncation orders $M$
to numerical simulations of AOUPs, RTPs, and ABPs in $d = 1$ and $2$,
showing good agreement.
The analytical solution captures the well-known accumulation of active particles at boundaries \cite{Elgeti2013was, Maggi2015msp, Solon2015abp, Bechinger2016apc, Das2018cab, Caprini2018apc, Sindelka2025cap}.
We highlight that the result for $M = 1$ is in fact exact for RTPs in $d = 1$ (cf.\ Fig.~\ref{fig:PositionDistribution}b),
as detailed in Section~\sref{app:ExactSolutionRTP} of~\cite{suppl_arxiv}.
For AOUPs (Fig.~\ref{fig:PositionDistribution}a),
the approximation improves with increasing truncation order as expected,
but the result for $M = 1$ [Eq.~\eqref{eq:rho:x:M1}] already provides a good representation,
especially for lower activity.
Remarkably, these approximations also align well with the numerical data for ABPs and RTPs in $d = 2$ (Figs.~\ref{fig:PositionDistribution}c and d),
even though the expansion does not systematically solve the corresponding Fokker-Planck equation exactly in these cases.

\paragraph*{Ideal gas law.}
Let us now return to dimensionful quantities $t$, $x$, $\eta$, etc., and denote the corresponding dimensionless versions explicitly as $\tilde t$, $\tilde x$, $\tilde\eta$, and so on.
To obtain an ideal gas law for our active particle systems,
we need to calculate the pressure $P$,
defined as the force per unit area that the particles exert on the walls of the enclosing box.
In our present setup of pointlike particles and rectangular walls,
we can exploit that the pressure is a state function \cite{Takatori:2014sps, Solon2015ppe, Winkler2015vps, Solon2015pin, Nikola2016aps, Digregorio2018fpd},
and thus
can be calculated from the virials of the active and thermal ``forces'' driving the particles [cf.\ Eq.~\eqref{eq:EOM:x}] \cite{Takatori:2014sps, Winkler2015vps, Speck2016sta, Speck2016ibp, Levis2017abe, Digregorio2018fpd, Note:ReferencePressure},
\begin{equation}
\label{eq:pressure}
	P = \frac{N \kB T}{V} + \frac{\gamma N}{V d} \< \bm x \cdot \bm v \> \,.
\end{equation}
The first term is the ordinary ideal-gas pressure for passive particles as it arises from the virial of the thermal fluctuations.
The second term is the active or swim pressure attributed to self-propulsion \cite{Takatori:2014sps, Yan2015sfa, Solon2015ppe, Winkler2015vps, Bialke:2015nit, Speck2016sta, Speck2016ibp, Junot2017avp, Levis2017abe, Digregorio2018fpd, Speck2020cfs}.

For AOUPs and a (hyper)cubic box,
we can adopt our steady-state solution to calculate the correlation function
$\< \bm x \cdot \bm v \> =  D \< \tilde{\bm x} \cdot \tilde{\bm v} \> = d \, D \, \Pe \, \< \tilde x \tilde\eta \>$.
Exploiting again the orthogonality property~\eqref{eq:He:Orthogonality} and $\He_1(\tilde\eta) = \tilde\eta$,
we find
\begin{equation}
\label{eq:xeta}
    \< \tilde x \tilde\eta \> = \int_0^{\tilde L} \!\!\! \d\tilde x \int_{-\infty}^\infty  \!\!\! \d\tilde\eta \, \rho(\tilde x, \tilde\eta) \tilde x \tilde\eta
        = \int_0^{\tilde L}  \!\!\! \d\tilde x \, X_1(\tilde x) \tilde x \,.
\end{equation}
At truncation order $M = 1$,
this becomes $\< \tilde x \tilde\eta \> = \Pe \, (1 - \vartheta) / (1 + \Pe^2 \vartheta)$ \cite{suppl_arxiv}.
Substituting into Eq.~\eqref{eq:pressure} and rearranging terms,
we then recover the ideal active gas law~\eqref{eq:IdealActiveGasLaw}.
As mentioned above, this result turns out to be exact for RTPs in $d = 1$
(see Section~\sref{app:ExactSolutionRTP} of~\cite{suppl_arxiv}).

\begin{figure}
\includegraphics[width=\columnwidth]{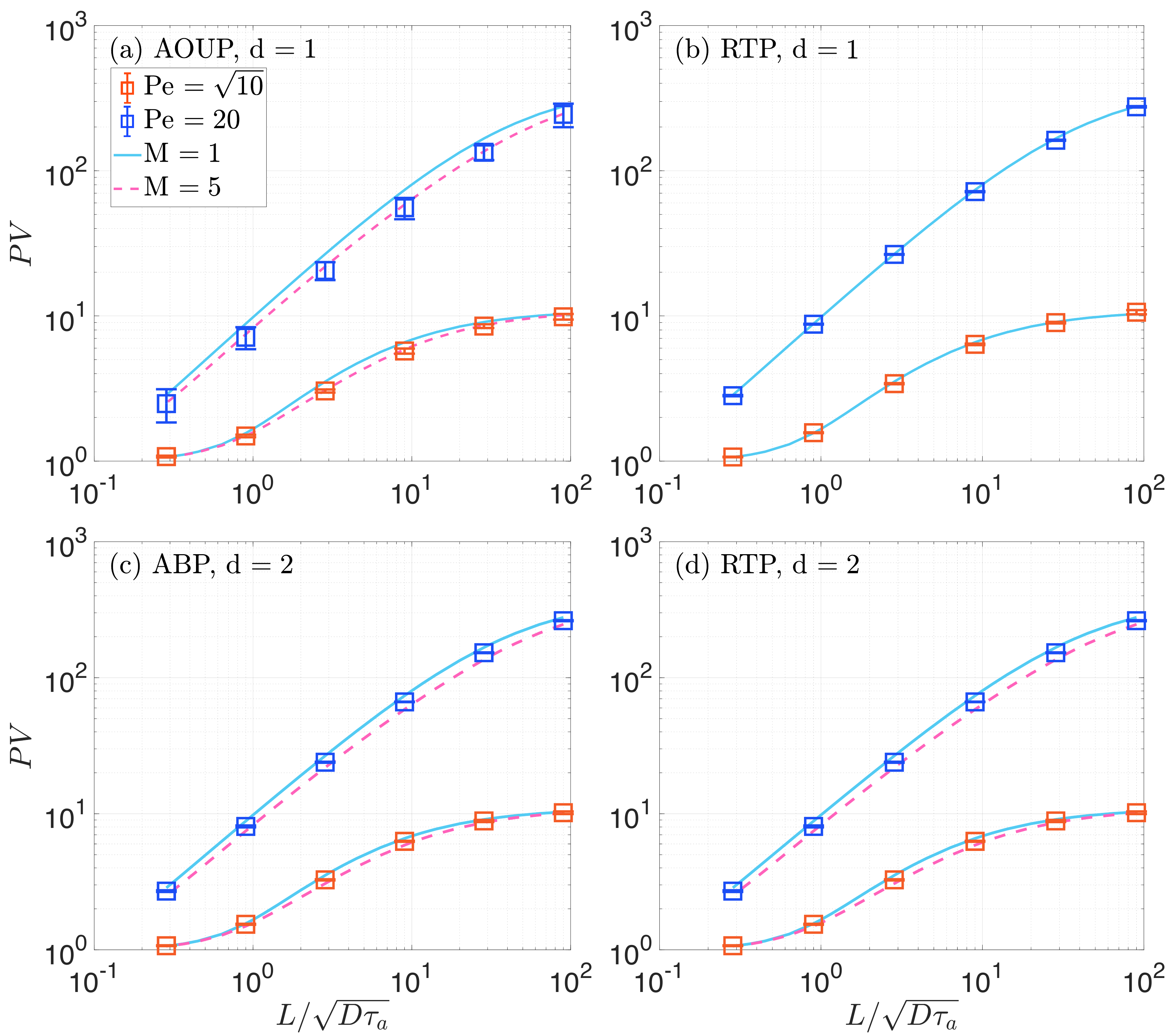}
\caption{Numerical verification of the ideal active gas law for different active particle models, system sizes, and activities.
Models and parameters in panels (a) to (d) are as in Fig.~\ref{fig:PositionDistribution},
see also legend.
Symbols show the numerically evaluated product $PV$ following Eq.~\eqref{eq:pressure},
i.e., the correlator $\< \bm x \cdot \bm v \>$ was calculated from simulations.
Solid lines correspond to the theoretical prediction from the right-hand side of Eq.~\eqref{eq:IdealActiveGasLaw},
i.e., the exact result for RTPs in $d = 1$ and the $M = 1$ approximation for AOUPs.
Dashed lines correspond to the $M = 5$ approximation.
}
\label{fig:IdealActiveGasLaw}
\end{figure}

In Fig.~\ref{fig:IdealActiveGasLaw},
we compare the analytical result with numerical simulations.
The solid lines show the right-hand side of Eq.~\eqref{eq:IdealActiveGasLaw},
while the dashed lines are the corresponding AOUP approximations of truncation order $M = 5$.
The numerical data (symbols) are calculated as $P V$ according to Eq.~\eqref{eq:pressure} from simulations of the $\< \bm x \cdot \bm v \>$ correlations (see also Section~\sref{app:Numerics} of~\cite{suppl_arxiv}).
Agreement is very good and improves with the truncation order for AOUPs (Fig.~\ref{fig:IdealActiveGasLaw}a).
Fig.~\ref{fig:IdealActiveGasLaw}b confirms that Eq.~\eqref{eq:IdealActiveGasLaw} is exact for RTPs in $d = 1$,
while Figs.~\ref{fig:IdealActiveGasLaw}c and d demonstrate that this relation and its $M = 5$ analog remain useful approximations for ABPs and RTPs in higher dimensions.

If the linear dimensions of the box are not equal and instead given by $L_1, \ldots, L_d$,
the pressure becomes anisotropic.
In the AOUP model, the
dynamics in the $d$ spatial dimensions are
independent,
so the pressure exerted on the walls in the $k$th dimension is
\begin{equation}
\label{eq:Pk}
    P_k = \frac{N \kB T}{V} \left( \frac{1 + \Pe^2}{1 + \Pe^2 \, \vartheta(L_k)} \right) ,
\end{equation}
where $\vartheta(L) := \frac{\tanh(\ell(L))}{\ell(L)}$ and $\ell(L) := L \sqrt{\frac{1 + \Pe^2}{4 D \ta}}$ as before.
The total pressure is the average over all dimensions, $P = \frac{1}{d} \sum_{k=1}^d P_k$,
and satisfies the generalized ideal active gas law
\begin{equation}
\label{eq:IdealActiveGasLaw:multi}
    P V = N \kB T \frac{1}{d} \, \sum_{k=1}^d \frac{1 + \Pe^2}{1 + \Pe^2 \vartheta(L_k)} \,.
\end{equation}

Likewise, higher-order truncations of the AOUP solution still yield a law of the form~\eqref{eq:IdealActiveGasLaw:multi} [or~\eqref{eq:IdealActiveGasLaw} for hypercubic boxes],
but with modified
$\vartheta = \vartheta(L)$, namely
\begin{equation}
\label{eq:IdealActiveGasLaw:HigherOrderTheta}
    \vartheta = \sum_{m=1}^M w_m \frac{\tanh \ell^{(m)}}{\ell^{(m)}} \,,
\end{equation}
where the weights $w_m$ satisfy $\sum_{m=1}^M w_m = 1$
and the $\ell^{(m)}$ are effective dimensionless length scales;
see Section~\sref{app:HigherOrderAOUPLaw} of~\cite{suppl_arxiv} for details.

\begin{figure*}
\centering
\includegraphics[width=2\columnwidth]{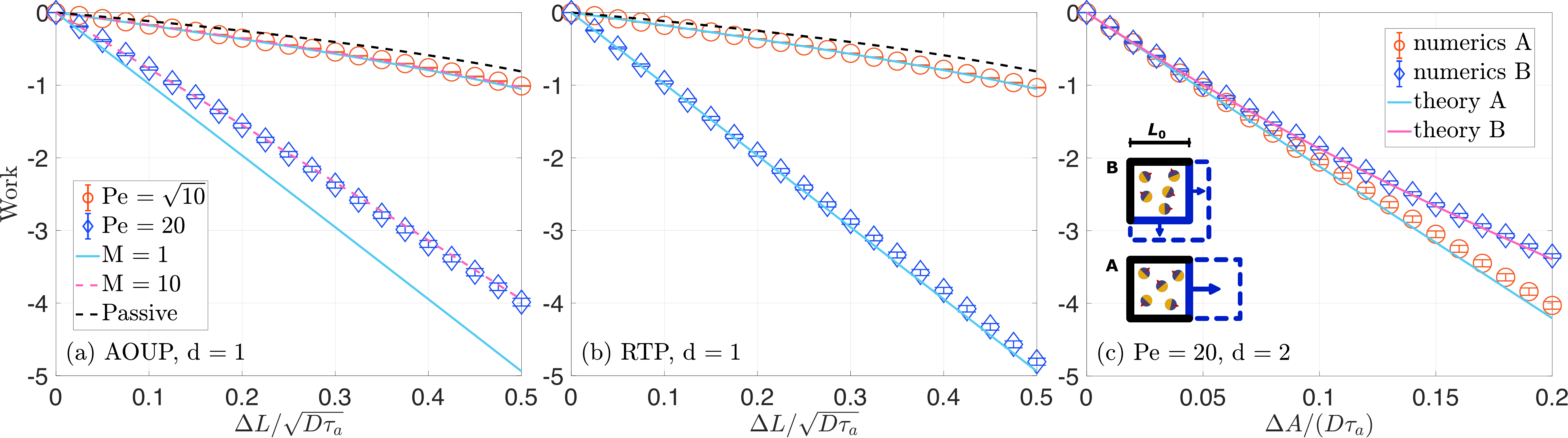}
\caption{
Work generated during expansion processes of ideal active gases.
(a) AOUPs in $d = 1$, linear expansion from $L_{\initial} = 0.4$ to $L_{\final} = L_{\initial} + \Delta L$. The total expansion time
for the largest shown volume change $\Delta L/\sqrt{D\ta}=0.5$
is $\tau = 70 \ta$ for $\Pe = \sqrt{10}$ and $\tau = 20 \ta$ for $\Pe = 20$.
Symbols correspond to numerical results from Langevin simulations using standard stochastic energetics for a wall potential [Eq.~\eqref{eq:WallPotential}].
Lines represent the theoretical work~\eqref{eq:W} for adiabatic expansion calculated from the ideal active gas law~\eqref{eq:IdealActiveGasLaw}
($M=1$) and its higher-order generalizations.
The black dashed line shows the passive limit $\Pe=0$.
(b) Same as in (a), but for RTPs in $d=1$ and exploiting the exact law~\eqref{eq:IdealActiveGasLaw}.
(c) AOUPs in $d = 2$ for two different protocols, \eqref{eq:Protocol:LA} and~\eqref{eq:Protocol:LB}, with $\tau = 20 \ta$ and $\Pe = 20$.
The time-dependent linear boxes sizes $L_1(t)$ and $L_2(t)$ are different in the two protocols, but the total volume/area $V(t) = L_1(t) L_2(t)$ is the same,
cf.\ the sketch in the inset.
Symbols (numerics) are obtained as in (a) and (b),
lines (theory) correspond to~\eqref{eq:W} using the $M = 5$ approximation to calculate $P_k$.
}
\label{fig:Process}
\end{figure*}

\paragraph*{Thermodynamic processes.}
One virtue of relations such as Eq.~\eqref{eq:IdealActiveGasLaw} is that they allow us to describe thermodynamic processes involving active matter at a macroscopic level,
without recourse to particle trajectories and the full machinery of stochastic thermodynamics \cite{Sekimoto2010se, Seifert2012stf, Peliti2021sti, Shiraishi2023ist}.
These processes have to be carried out sufficiently slowly since Eq.~\eqref{eq:IdealActiveGasLaw} is based on a stationarity assumption,
just like the original ideal gas law for passive particles.

For example,
we can analyze
the work needed to compress an active matter system
or the work gained upon expanding it.
The time-dependent linear dimensions of the box $\bm L(t) = [L_1(t), \ldots, L_d(t)]$ for $t \in [0, \tau]$ define the protocol of
the compression or expansion process.
Then the time-dependent box volume is
$V(t) = \prod_{k=1}^d L_k(t)$, and
$\d V = \sum_{k=1}^d \d V_k$ is the infinitesimal volume element,
with $\d V_k = \frac{\d L_k}{L_k} V$.
The thermodynamic work required to carry out this protocol is
\begin{equation}
\label{eq:W}
\begin{aligned}
    W &= \sum_{k=1}^d \int \d V_k \, P_k \\
        &= N \kB T \sum_{k=1}^d \int_{L_k(0)}^{L_k(\tau)} \! \frac{\d L_k}{L_k} \frac{1 + \Pe^2}{1 + \Pe^2 \vartheta(L_k)} \,.
\end{aligned}
\end{equation}
To demonstrate the usefulness and practicality of this macroscopic approach,
we compare the
work $W$, Eq.~\eqref{eq:W}, for AOUPs and RTPs
with a microscopic numerical simulation and evaluation using the framework of stochastic thermodynamics \cite{Jarzynski2011eii, Seifert2012stf, Peliti2021sti, Shiraishi2023ist}.
To this end,
we model the walls by a steep potential
\begin{subequations}
\label{eq:WallPotential}
\begin{align}
    U(x, L) &:= \Theta(-x) \Psi(x) + \Theta(x-L) \Psi(L-x) \,, \\
    \Psi(x) &:= \left[ \left(\frac{\sigma}{\sigma + x}\right)^\nu - 1 \right]^2
\end{align}
\end{subequations}
in every dimension,
where $\Theta(x)$ is the Heaviside step function
and $\sigma = 0.05$ and $\nu = 4$.
The work is obtained from averages over microscopic trajectories as
\begin{equation}
\label{eq:W:micro}
    W = \sum_{k=1}^d \int_0^{\tau} \d t \, \dot{L}_k(t) \left\< \frac{\partial U}{\partial L_k}(x_k(t), L_k(t)) \right\> .
\end{equation}
Further details of the numerical simulations are collected in Section~\sref{app:Numerics} of~\cite{suppl_arxiv}.

A comparison between our macroscopic theory and the microscopic numerics is presented for different expansion processes of a 1D gas in Fig.~\ref{fig:Process}a (AOUPs) and b (RTPs).
The $M=10$ approximation yields excellent agreement for AOUPs in Fig.~\ref{fig:Process}a,
but even the lowest order approximation ($M = 1$ and Eq.~\eqref{eq:IdealActiveGasLaw}) provides a decent estimate of the extractable work from this process,
especially if the activity is not too high.
For RTPs in Fig.~\ref{fig:Process}b,
we recall that the theory is exact for hard walls in the limit of slow processes,
hence the small deviations can be attributed entirely to finite process times and the ``softness'' of the wall, being modeled by a potential in the numerics.

Remarkably, as a genuine feature of active matter that cannot be found in passive systems,
Eqs.~\eqref{eq:IdealActiveGasLaw:multi} and~\eqref{eq:W} predict that the work needed or gained from the volume change of an active gas depends on the \emph{shape} of the volume in dimensions $d>1$.
To illustrate this, we consider two protocols in $d = 2$ which yield the same time-dependent total volume $V(t)$,
but have different linear deformations;
see the inset of Fig.~\ref{fig:Process}c for an illustration.
In the first protocol, $\bm L^{\mathrm{A}}(t)$,
we only change the box length in one direction while keeping the other direction fixed,
\begin{equation}
\label{eq:Protocol:LA}
    L^{\mathrm{A}}_1(t) = L_{\initial} + (L_{\final} - L_{\initial}) \frac{t}{\tau} \,,
    \quad
    L^{\mathrm{A}}_2(t) = L_0 = \mathrm{const} \,.
\end{equation}
In the second protocol, $\bm L^{\mathrm{B}}(t)$,
we change the box length in both directions in the same way,
\begin{equation}
\label{eq:Protocol:LB}
    L^{\mathrm{B}}_1(t) = L^{\mathrm{B}}_2(t) = \sqrt{ L_0 [ L_{\initial} + (L_{\final} - L_{\initial}) \frac{t}{\tau} ] } \,.
\end{equation}
In a passive system, the thermodynamic work of processes A and B would be identical.
Likewise, the law for the thermodynamic limit, $PV = N \kB T (1 + \Pe^2)$ \cite{Solon2015pin, Solon2015ppe, Winkler2015vps, MariniBettoloMarconi2015tsm, Cameron2023esa, Dabelow2025tni} would yield the same work in both cases.
The numerics in Fig.~\ref{fig:Process}c,
however,
confirms the prediction of our finite-volume ideal active gas law,
cf.\ Eqs.~\eqref{eq:Pk} and~\eqref{eq:W}:
It demonstrates that
the energetics of thermodynamic processes in active systems depend on the box sizes in every dimension.
The work that can be extracted from an expansion process is generally higher under protocol A than under protocol B.
Moreover, our theory reproduces the numerically measured expansion work very well quantitatively again,
based on macroscopic quantities alone,
without resorting to the microscopic particle trajectories.

\paragraph*{Conclusions.}
Based on steady-state solutions for noninteracting active particles confined by hard walls,
we derived an ideal gas law for active matter, Eq.~\eqref{eq:IdealActiveGasLaw}.
The law is exact for RTPs in $d=1$, corresponds to the first order of a systematic expansion for AOUPs in hypercubic boxes in arbitrary dimensions, and was found numerically to provide a good approximation for RTPs and ABPs in higher dimensions.
It therefore captures the behavior of the three standard minimal models for active matter at mesoscopic scales (AOUPs, RTPs, and ABPs) in good approximation,
in line with similar large-scale phenomenology observed among them \cite{Fily2012aps, Cates2013waa, Farage2015eia, Solon2015abp, Caprini2019cst}.
For other rectangular geometries, the law can be extended into Eq.~\eqref{eq:IdealActiveGasLaw:multi} and correctly predicts a shape dependence of thermodynamic properties absent in passive systems.

It
generalizes
several important, previously known special cases.
As $\Pe \to 0$, $\ta \to \infty$, or $L \to 0$ (at fixed $N/V$),
the systems behave like a passive ideal gas at temperature $T = \gamma D / \kB$.
If $\ta \to 0$ or $L \to \infty$ (at fixed $N/V$, thermodynamic limit),
they also become effectively passive,
but at a higher effective temperature $T_{\text{eff}} = T + \gamma \Da / \kB$.
These limits agree for AOUPs, RTPs, and ABPs \cite{Solon2015pin, Solon2015ppe, Winkler2015vps, Cameron2023esa, MariniBettoloMarconi2015tsm, Dabelow2025tni}.
Notably, the relation for the large system size limit is valid only if \emph{all} linear box dimensions satisfy $L_k \gg \sqrt{\frac{4 D \ta}{1 + \Pe^2}}$ [see the definitions of $\vartheta$ and $\ell$ below Eq.~\eqref{eq:Pk}],
whereas our generalization covers the full range of $L_k$ values.

An important application of the law~\eqref{eq:IdealActiveGasLaw} and its extensions is that it allows us to describe thermodynamic processes with active matter from a macroscopic perspective.
Stochastic thermodynamics is an extremely successful framework that generalizes macroscopic thermodynamics to mesoscopic systems,
which are strongly affected by thermal fluctuations.
However, evaluating thermodynamic quantities such as work and heat within this framework in specific systems typically requires knowledge of individual particle trajectories
in order to integrate the defining stochastic functionals.
By contrast, the ideal active gas law provides access to thermodynamic quantities through macroscopic state variables such as the pressure, volume, and temperature alone,
which we illustrated by analyzing the work in various expansion processes.
This macroscopic thermodynamic analysis relies on our law holding
for finite system sizes. In particular, it accurately captures the non-trivial interplay between the persistence of active motion and the finite box dimensions, which does not exist in passive systems or the large-system limit.
This paves the way towards a macroscopic description of thermodynamic cycles and engines involving strongly confined active matter \cite{Krishnamurthy2016msh, Martin2018emp, Pietzonka2019aed, Ekeh2020tca, Datta2022sla, Davis2024amc, GarciaMillan2025ocl, Davis2026omp}.

It should be pointed out that there are limitations to treating general active matter systems with the law found here.
In particular,
it requires sufficiently slow processes and weak interactions or low densities.
However, such restrictions equally apply to the passive ideal gas law,
whose importance in science and engineering applications is undoubted.
Finding suitable generalizations that take other aspects of real active matter systems into account will be an important future direction.
Besides interactions and hydrodynamic effects, these include complications specific to active matter, arising from torques due to nonspherical particle shapes or walls with non-constant curvature \cite{Solon2015pin, Nikola2016aps, Junot2017avp}
and the intricacies of modeling the chemical processes underlying self-propulsion \cite{Gaspard2018fcs, Pietzonka2018epa, Gaspard2019tsm, Fritz2023tcm, Chatzittofi2024ept}, for example.

\begin{acknowledgments}
We thank Jaime Agudo-Canalejo and Rainer Klages for insightful comments.
SL acknowledges support from the China Scholarship Council (CSC).
This research utilized Queen Mary's Apocrita HPC facility, supported by QMUL Research-IT \cite{King2017ahp}.
RE acknowledges funding by the Swedish Research Council (Vetenskapsrådet) under Grant No.~2024-05091.
\end{acknowledgments}


\begin{thebibliography}{61}%
\makeatletter
\providecommand \@ifxundefined [1]{%
 \@ifx{#1\undefined}
}%
\providecommand \@ifnum [1]{%
 \ifnum #1\expandafter \@firstoftwo
 \else \expandafter \@secondoftwo
 \fi
}%
\providecommand \@ifx [1]{%
 \ifx #1\expandafter \@firstoftwo
 \else \expandafter \@secondoftwo
 \fi
}%
\providecommand \natexlab [1]{#1}%
\providecommand \enquote  [1]{``#1''}%
\providecommand \bibnamefont  [1]{#1}%
\providecommand \bibfnamefont [1]{#1}%
\providecommand \citenamefont [1]{#1}%
\providecommand \href@noop [0]{\@secondoftwo}%
\providecommand \href [0]{\begingroup \@sanitize@url \@href}%
\providecommand \@href[1]{\@@startlink{#1}\@@href}%
\providecommand \@@href[1]{\endgroup#1\@@endlink}%
\providecommand \@sanitize@url [0]{\catcode `\\12\catcode `\$12\catcode
  `\&12\catcode `\#12\catcode `\^12\catcode `\_12\catcode `\%12\relax}%
\providecommand \@@startlink[1]{}%
\providecommand \@@endlink[0]{}%
\providecommand \url  [0]{\begingroup\@sanitize@url \@url }%
\providecommand \@url [1]{\endgroup\@href {#1}{\urlprefix }}%
\providecommand \urlprefix  [0]{URL }%
\providecommand \Eprint [0]{\href }%
\providecommand \doibase [0]{https://doi.org/}%
\providecommand \selectlanguage [0]{\@gobble}%
\providecommand \bibinfo  [0]{\@secondoftwo}%
\providecommand \bibfield  [0]{\@secondoftwo}%
\providecommand \translation [1]{[#1]}%
\providecommand \BibitemOpen [0]{}%
\providecommand \bibitemStop [0]{}%
\providecommand \bibitemNoStop [0]{.\EOS\space}%
\providecommand \EOS [0]{\spacefactor3000\relax}%
\providecommand \BibitemShut  [1]{\csname bibitem#1\endcsname}%
\let\auto@bib@innerbib\@empty
\bibitem [{\citenamefont {Landau}\ and\ \citenamefont
  {Lifshitz}(1970)}]{LandauLifshitz:StatisticalPhysics}%
  \BibitemOpen
  \bibfield  {author} {\bibinfo {author} {\bibfnamefont {L.}~\bibnamefont
  {Landau}}\ and\ \bibinfo {author} {\bibfnamefont {E.}~\bibnamefont
  {Lifshitz}},\ }\href@noop {} {\emph {\bibinfo {title} {Statistical
  Physics}}}\ (\bibinfo  {publisher} {Pergamon, Oxford},\ \bibinfo {year}
  {1970})\BibitemShut {NoStop}%
\bibitem [{\citenamefont {Kittel}\ and\ \citenamefont
  {Kroemer}(1980)}]{KittelKroemer:ThermalPhysics}%
  \BibitemOpen
  \bibfield  {author} {\bibinfo {author} {\bibfnamefont {C.}~\bibnamefont
  {Kittel}}\ and\ \bibinfo {author} {\bibfnamefont {H.}~\bibnamefont
  {Kroemer}},\ }\href@noop {} {\emph {\bibinfo {title} {Thermal Physics}}},\
  \bibinfo {edition} {2nd}\ ed.\ (\bibinfo  {publisher} {W. H. Freeman},\
  \bibinfo {address} {US},\ \bibinfo {year} {1980})\BibitemShut {NoStop}%
\bibitem [{\citenamefont {Fodor}\ \emph {et~al.}(2022)\citenamefont {Fodor},
  \citenamefont {Jack},\ and\ \citenamefont {Cates}}]{Fodor:2022ibe}%
  \BibitemOpen
  \bibfield  {author} {\bibinfo {author} {\bibfnamefont {E.}~\bibnamefont
  {Fodor}}, \bibinfo {author} {\bibfnamefont {R.~L.}\ \bibnamefont {Jack}},\
  and\ \bibinfo {author} {\bibfnamefont {M.~E.}\ \bibnamefont {Cates}},\
  }\bibfield  {title} {\bibinfo {title} {Irreversibility and biased ensembles
  in active matter: Insights from stochastic thermodynamics},\ }\href
  {https://doi.org/10.1146/annurev-conmatphys-031720-032419} {\bibfield
  {journal} {\bibinfo  {journal} {Annual Review of Condensed Matter Physics}\
  }\textbf {\bibinfo {volume} {13}},\ \bibinfo {pages} {215} (\bibinfo {year}
  {2022})},\ \Eprint
  {https://arxiv.org/abs/https://doi.org/10.1146/annurev-conmatphys-031720-032419}
  {https://doi.org/10.1146/annurev-conmatphys-031720-032419} \BibitemShut
  {NoStop}%
\bibitem [{\citenamefont {Romanczuk}\ \emph {et~al.}(2012)\citenamefont
  {Romanczuk}, \citenamefont {B{\"a}r}, \citenamefont {Ebeling}, \citenamefont
  {Lindner},\ and\ \citenamefont {Schimansky-Geier}}]{Romanczuk2012abp}%
  \BibitemOpen
  \bibfield  {author} {\bibinfo {author} {\bibfnamefont {P.}~\bibnamefont
  {Romanczuk}}, \bibinfo {author} {\bibfnamefont {M.}~\bibnamefont {B{\"a}r}},
  \bibinfo {author} {\bibfnamefont {W.}~\bibnamefont {Ebeling}}, \bibinfo
  {author} {\bibfnamefont {B.}~\bibnamefont {Lindner}},\ and\ \bibinfo {author}
  {\bibfnamefont {L.}~\bibnamefont {Schimansky-Geier}},\ }\bibfield  {title}
  {\bibinfo {title} {Active brownian particles},\ }\href
  {https://doi.org/10.1140/epjst/e2012-01529-y} {\bibfield  {journal} {\bibinfo
   {journal} {The European Physical Journal Special Topics}\ }\textbf {\bibinfo
  {volume} {202}},\ \bibinfo {pages} {1} (\bibinfo {year} {2012})}\BibitemShut
  {NoStop}%
\bibitem [{\citenamefont {Bechinger}\ \emph {et~al.}(2016)\citenamefont
  {Bechinger}, \citenamefont {Di~Leonardo}, \citenamefont {L\"owen},
  \citenamefont {Reichhardt}, \citenamefont {Volpe},\ and\ \citenamefont
  {Volpe}}]{Bechinger2016apc}%
  \BibitemOpen
  \bibfield  {author} {\bibinfo {author} {\bibfnamefont {C.}~\bibnamefont
  {Bechinger}}, \bibinfo {author} {\bibfnamefont {R.}~\bibnamefont
  {Di~Leonardo}}, \bibinfo {author} {\bibfnamefont {H.}~\bibnamefont
  {L\"owen}}, \bibinfo {author} {\bibfnamefont {C.}~\bibnamefont {Reichhardt}},
  \bibinfo {author} {\bibfnamefont {G.}~\bibnamefont {Volpe}},\ and\ \bibinfo
  {author} {\bibfnamefont {G.}~\bibnamefont {Volpe}},\ }\bibfield  {title}
  {\bibinfo {title} {Active particles in complex and crowded environments},\
  }\href {https://doi.org/10.1103/RevModPhys.88.045006} {\bibfield  {journal}
  {\bibinfo  {journal} {Rev. Mod. Phys.}\ }\textbf {\bibinfo {volume} {88}},\
  \bibinfo {pages} {045006} (\bibinfo {year} {2016})}\BibitemShut {NoStop}%
\bibitem [{\citenamefont {Takatori}\ and\ \citenamefont
  {Brady}(2016)}]{Takatori2016fst}%
  \BibitemOpen
  \bibfield  {author} {\bibinfo {author} {\bibfnamefont {S.~C.}\ \bibnamefont
  {Takatori}}\ and\ \bibinfo {author} {\bibfnamefont {J.~F.}\ \bibnamefont
  {Brady}},\ }\bibfield  {title} {\bibinfo {title} {Forces, stresses and the
  (thermo?) dynamics of active matter},\ }\href
  {https://doi.org/https://doi.org/10.1016/j.cocis.2015.12.003} {\bibfield
  {journal} {\bibinfo  {journal} {Curr. Opin. Colloid In.}\ }\textbf {\bibinfo
  {volume} {21}},\ \bibinfo {pages} {24} (\bibinfo {year} {2016})}\BibitemShut
  {NoStop}%
\bibitem [{\citenamefont {Solon}\ \emph
  {et~al.}(2015{\natexlab{a}})\citenamefont {Solon}, \citenamefont {Fily},
  \citenamefont {Baskaran}, \citenamefont {Cates}, \citenamefont {Kafri},
  \citenamefont {Kardar},\ and\ \citenamefont {Tailleur}}]{Solon2015pin}%
  \BibitemOpen
  \bibfield  {author} {\bibinfo {author} {\bibfnamefont {A.~P.}\ \bibnamefont
  {Solon}}, \bibinfo {author} {\bibfnamefont {Y.}~\bibnamefont {Fily}},
  \bibinfo {author} {\bibfnamefont {A.}~\bibnamefont {Baskaran}}, \bibinfo
  {author} {\bibfnamefont {M.~E.}\ \bibnamefont {Cates}}, \bibinfo {author}
  {\bibfnamefont {Y.}~\bibnamefont {Kafri}}, \bibinfo {author} {\bibfnamefont
  {M.}~\bibnamefont {Kardar}},\ and\ \bibinfo {author} {\bibfnamefont
  {J.}~\bibnamefont {Tailleur}},\ }\bibfield  {title} {\bibinfo {title}
  {Pressure is not a state function for generic active fluids},\ }\href
  {https://doi.org/10.1038/nphys3377} {\bibfield  {journal} {\bibinfo
  {journal} {Nature Physics}\ }\textbf {\bibinfo {volume} {11}},\ \bibinfo
  {pages} {673} (\bibinfo {year} {2015}{\natexlab{a}})}\BibitemShut {NoStop}%
\bibitem [{\citenamefont {Solon}\ \emph
  {et~al.}(2015{\natexlab{b}})\citenamefont {Solon}, \citenamefont
  {Stenhammar}, \citenamefont {Wittkowski}, \citenamefont {Kardar},
  \citenamefont {Kafri}, \citenamefont {Cates},\ and\ \citenamefont
  {Tailleur}}]{Solon2015ppe}%
  \BibitemOpen
  \bibfield  {author} {\bibinfo {author} {\bibfnamefont {A.~P.}\ \bibnamefont
  {Solon}}, \bibinfo {author} {\bibfnamefont {J.}~\bibnamefont {Stenhammar}},
  \bibinfo {author} {\bibfnamefont {R.}~\bibnamefont {Wittkowski}}, \bibinfo
  {author} {\bibfnamefont {M.}~\bibnamefont {Kardar}}, \bibinfo {author}
  {\bibfnamefont {Y.}~\bibnamefont {Kafri}}, \bibinfo {author} {\bibfnamefont
  {M.~E.}\ \bibnamefont {Cates}},\ and\ \bibinfo {author} {\bibfnamefont
  {J.}~\bibnamefont {Tailleur}},\ }\bibfield  {title} {\bibinfo {title}
  {Pressure and phase equilibria in interacting active brownian spheres},\
  }\href {https://doi.org/10.1103/PhysRevLett.114.198301} {\bibfield  {journal}
  {\bibinfo  {journal} {Phys. Rev. Lett.}\ }\textbf {\bibinfo {volume} {114}},\
  \bibinfo {pages} {198301} (\bibinfo {year} {2015}{\natexlab{b}})}\BibitemShut
  {NoStop}%
\bibitem [{\citenamefont {Winkler}\ \emph {et~al.}(2015)\citenamefont
  {Winkler}, \citenamefont {Wysocki},\ and\ \citenamefont
  {Gompper}}]{Winkler2015vps}%
  \BibitemOpen
  \bibfield  {author} {\bibinfo {author} {\bibfnamefont {R.~G.}\ \bibnamefont
  {Winkler}}, \bibinfo {author} {\bibfnamefont {A.}~\bibnamefont {Wysocki}},\
  and\ \bibinfo {author} {\bibfnamefont {G.}~\bibnamefont {Gompper}},\
  }\bibfield  {title} {\bibinfo {title} {Virial pressure in systems of
  spherical active brownian particles},\ }\href
  {https://doi.org/10.1039/C5SM01412C} {\bibfield  {journal} {\bibinfo
  {journal} {Soft Matter}\ }\textbf {\bibinfo {volume} {11}},\ \bibinfo {pages}
  {6680} (\bibinfo {year} {2015})}\BibitemShut {NoStop}%
\bibitem [{\citenamefont {Marini Bettolo~Marconi}\ and\ \citenamefont
  {Maggi}(2015)}]{MariniBettoloMarconi2015tsm}%
  \BibitemOpen
  \bibfield  {author} {\bibinfo {author} {\bibfnamefont {U.}~\bibnamefont
  {Marini Bettolo~Marconi}}\ and\ \bibinfo {author} {\bibfnamefont
  {C.}~\bibnamefont {Maggi}},\ }\bibfield  {title} {\bibinfo {title} {Towards a
  statistical mechanical theory of active fluids},\ }\href
  {https://doi.org/10.1039/C5SM01718A} {\bibfield  {journal} {\bibinfo
  {journal} {Soft Matter}\ }\textbf {\bibinfo {volume} {11}},\ \bibinfo {pages}
  {8768} (\bibinfo {year} {2015})}\BibitemShut {NoStop}%
\bibitem [{\citenamefont {Cameron}\ \emph {et~al.}(2023)\citenamefont
  {Cameron}, \citenamefont {Mosayebi}, \citenamefont {Bennett},\ and\
  \citenamefont {Liverpool}}]{Cameron2023esa}%
  \BibitemOpen
  \bibfield  {author} {\bibinfo {author} {\bibfnamefont {S.}~\bibnamefont
  {Cameron}}, \bibinfo {author} {\bibfnamefont {M.}~\bibnamefont {Mosayebi}},
  \bibinfo {author} {\bibfnamefont {R.}~\bibnamefont {Bennett}},\ and\ \bibinfo
  {author} {\bibfnamefont {T.~B.}\ \bibnamefont {Liverpool}},\ }\bibfield
  {title} {\bibinfo {title} {Equation of state for active matter},\ }\href
  {https://doi.org/10.1103/PhysRevE.108.014608} {\bibfield  {journal} {\bibinfo
   {journal} {Phys. Rev. E}\ }\textbf {\bibinfo {volume} {108}},\ \bibinfo
  {pages} {014608} (\bibinfo {year} {2023})}\BibitemShut {NoStop}%
\bibitem [{\citenamefont {Dabelow}\ and\ \citenamefont
  {Eichhorn}(2025)}]{Dabelow2025tni}%
  \BibitemOpen
  \bibfield  {author} {\bibinfo {author} {\bibfnamefont {L.}~\bibnamefont
  {Dabelow}}\ and\ \bibinfo {author} {\bibfnamefont {R.}~\bibnamefont
  {Eichhorn}},\ }\bibfield  {title} {\bibinfo {title} {Thermodynamic nature of
  irreversibility in active matter},\ }\href
  {https://doi.org/10.1103/gbzm-g8ky} {\bibfield  {journal} {\bibinfo
  {journal} {Phys. Rev. Res.}\ }\textbf {\bibinfo {volume} {7}},\ \bibinfo
  {pages} {033077} (\bibinfo {year} {2025})}\BibitemShut {NoStop}%
\bibitem [{\citenamefont {Risken}(1984)}]{Risken1984fpe}%
  \BibitemOpen
  \bibfield  {author} {\bibinfo {author} {\bibfnamefont {H.}~\bibnamefont
  {Risken}},\ }\href@noop {} {\emph {\bibinfo {title} {The Fokker-Planck
  Equation}}}\ (\bibinfo  {publisher} {Springer},\ \bibinfo {year}
  {1984})\BibitemShut {NoStop}%
\bibitem [{\citenamefont {Gardiner}(1985)}]{Gardiner1985hsm}%
  \BibitemOpen
  \bibfield  {author} {\bibinfo {author} {\bibfnamefont {C.}~\bibnamefont
  {Gardiner}},\ }\href@noop {} {\emph {\bibinfo {title} {Handbook of Stochastic
  Methods for Physics, Chemistry, and the Natural Sciences}}},\ \bibinfo
  {edition} {2nd}\ ed.\ (\bibinfo  {publisher} {Springer},\ \bibinfo {year}
  {1985})\ \bibinfo {note} {4th reprint, 1997}\BibitemShut {NoStop}%
\bibitem [{\citenamefont {Caprini}\ and\ \citenamefont {Marini
  Bettolo~Marconi}(2018)}]{Caprini2018apc}%
  \BibitemOpen
  \bibfield  {author} {\bibinfo {author} {\bibfnamefont {L.}~\bibnamefont
  {Caprini}}\ and\ \bibinfo {author} {\bibfnamefont {U.}~\bibnamefont {Marini
  Bettolo~Marconi}},\ }\bibfield  {title} {\bibinfo {title} {Active particles
  under confinement and effective force generation among surfaces},\ }\href
  {https://doi.org/10.1039/C8SM01840E} {\bibfield  {journal} {\bibinfo
  {journal} {Soft Matter}\ }\textbf {\bibinfo {volume} {14}},\ \bibinfo {pages}
  {9044} (\bibinfo {year} {2018})}\BibitemShut {NoStop}%
\bibitem [{\citenamefont {Garcia-Millan}\ and\ \citenamefont
  {Pruessner}(2021)}]{GarciaMillan2021rtm}%
  \BibitemOpen
  \bibfield  {author} {\bibinfo {author} {\bibfnamefont {R.}~\bibnamefont
  {Garcia-Millan}}\ and\ \bibinfo {author} {\bibfnamefont {G.}~\bibnamefont
  {Pruessner}},\ }\bibfield  {title} {\bibinfo {title} {Run-and-tumble motion
  in a harmonic potential: field theory and entropy production},\ }\href
  {https://doi.org/10.1088/1742-5468/ac014d} {\bibfield  {journal} {\bibinfo
  {journal} {J. Stat. Mech: Theory Exp.}\ }\textbf {\bibinfo {volume} {2021}},\
  \bibinfo {pages} {063203} (\bibinfo {year} {2021})}\BibitemShut {NoStop}%
\bibitem [{\citenamefont {Abramowitz}\ and\ \citenamefont
  {Stegun}(1972)}]{Abramowitz1972hmf}%
  \BibitemOpen
  \bibinfo {editor} {\bibfnamefont {M.}~\bibnamefont {Abramowitz}}\ and\
  \bibinfo {editor} {\bibfnamefont {I.~A.}\ \bibnamefont {Stegun}},\ eds.,\
  \href@noop {} {\emph {\bibinfo {title} {{Handbook of Mathematical Functions
  with Formulas, Graphs, and Mathematical Tables}}}}\ (\bibinfo  {publisher}
  {National Bureau of Standards},\ \bibinfo {address} {Washington, D.C.},\
  \bibinfo {year} {1972})\BibitemShut {NoStop}%
\bibitem [{sup()}]{suppl_arxiv}%
  \BibitemOpen
  \href@noop {} {}\bibinfo {note} {See Supplemental Material.}\BibitemShut
  {Stop}%
\bibitem [{\citenamefont {Elgeti}\ and\ \citenamefont
  {Gompper}(2013)}]{Elgeti2013was}%
  \BibitemOpen
  \bibfield  {author} {\bibinfo {author} {\bibfnamefont {J.}~\bibnamefont
  {Elgeti}}\ and\ \bibinfo {author} {\bibfnamefont {G.}~\bibnamefont
  {Gompper}},\ }\bibfield  {title} {\bibinfo {title} {Wall accumulation of
  self-propelled spheres},\ }\href
  {https://doi.org/10.1209/0295-5075/101/48003} {\bibfield  {journal} {\bibinfo
   {journal} {Europhys. Lett.}\ }\textbf {\bibinfo {volume} {101}},\ \bibinfo
  {pages} {48003} (\bibinfo {year} {2013})}\BibitemShut {NoStop}%
\bibitem [{\citenamefont {Maggi}\ \emph {et~al.}(2015)\citenamefont {Maggi},
  \citenamefont {Marconi}, \citenamefont {Gnan},\ and\ \citenamefont
  {Di~Leonardo}}]{Maggi2015msp}%
  \BibitemOpen
  \bibfield  {author} {\bibinfo {author} {\bibfnamefont {C.}~\bibnamefont
  {Maggi}}, \bibinfo {author} {\bibfnamefont {U.~M.~B.}\ \bibnamefont
  {Marconi}}, \bibinfo {author} {\bibfnamefont {N.}~\bibnamefont {Gnan}},\ and\
  \bibinfo {author} {\bibfnamefont {R.}~\bibnamefont {Di~Leonardo}},\
  }\bibfield  {title} {\bibinfo {title} {Multidimensional stationary
  probability distribution for interacting active particles},\ }\href
  {http://dx.doi.org/10.1038/srep10742} {\bibfield  {journal} {\bibinfo
  {journal} {Sci. Rep.}\ }\textbf {\bibinfo {volume} {5}},\ \bibinfo {pages}
  {10742} (\bibinfo {year} {2015})}\BibitemShut {NoStop}%
\bibitem [{\citenamefont {Solon}\ \emph
  {et~al.}(2015{\natexlab{c}})\citenamefont {Solon}, \citenamefont {Cates},\
  and\ \citenamefont {Tailleur}}]{Solon2015abp}%
  \BibitemOpen
  \bibfield  {author} {\bibinfo {author} {\bibfnamefont {A.~P.}\ \bibnamefont
  {Solon}}, \bibinfo {author} {\bibfnamefont {M.~E.}\ \bibnamefont {Cates}},\
  and\ \bibinfo {author} {\bibfnamefont {J.}~\bibnamefont {Tailleur}},\
  }\bibfield  {title} {\bibinfo {title} {Active brownian particles and
  run-and-tumble particles: A comparative study},\ }\href
  {https://doi.org/10.1140/epjst/e2015-02457-0} {\bibfield  {journal} {\bibinfo
   {journal} {Eur. Phys. J. Special Topics}\ }\textbf {\bibinfo {volume}
  {224}},\ \bibinfo {pages} {1231} (\bibinfo {year}
  {2015}{\natexlab{c}})}\BibitemShut {NoStop}%
\bibitem [{\citenamefont {Das}\ \emph {et~al.}(2018)\citenamefont {Das},
  \citenamefont {Gompper},\ and\ \citenamefont {Winkler}}]{Das2018cab}%
  \BibitemOpen
  \bibfield  {author} {\bibinfo {author} {\bibfnamefont {S.}~\bibnamefont
  {Das}}, \bibinfo {author} {\bibfnamefont {G.}~\bibnamefont {Gompper}},\ and\
  \bibinfo {author} {\bibfnamefont {R.~G.}\ \bibnamefont {Winkler}},\
  }\bibfield  {title} {\bibinfo {title} {Confined active brownian particles:
  theoretical description of propulsion-induced accumulation},\ }\href
  {https://doi.org/10.1088/1367-2630/aa9d4b} {\bibfield  {journal} {\bibinfo
  {journal} {New J. Phys.}\ }\textbf {\bibinfo {volume} {20}},\ \bibinfo
  {pages} {015001} (\bibinfo {year} {2018})}\BibitemShut {NoStop}%
\bibitem [{\citenamefont {Šindelka}\ \emph {et~al.}(2025)\citenamefont
  {Šindelka}, \citenamefont {Gadermeteva},\ and\ \citenamefont
  {Lísal}}]{Sindelka2025cap}%
  \BibitemOpen
  \bibfield  {author} {\bibinfo {author} {\bibfnamefont {K.}~\bibnamefont
  {Šindelka}}, \bibinfo {author} {\bibfnamefont {A.}~\bibnamefont
  {Gadermeteva}},\ and\ \bibinfo {author} {\bibfnamefont {M.}~\bibnamefont
  {Lísal}},\ }\bibfield  {title} {\bibinfo {title} {Confined active particles:
  wall accumulation and correspondence between active and fluid systems},\
  }\href {https://doi.org/10.1039/D5SM00495K} {\bibfield  {journal} {\bibinfo
  {journal} {Soft Matter}\ }\textbf {\bibinfo {volume} {21}},\ \bibinfo {pages}
  {7544} (\bibinfo {year} {2025})}\BibitemShut {NoStop}%
\bibitem [{\citenamefont {Takatori}\ \emph {et~al.}(2014)\citenamefont
  {Takatori}, \citenamefont {Yan},\ and\ \citenamefont
  {Brady}}]{Takatori:2014sps}%
  \BibitemOpen
  \bibfield  {author} {\bibinfo {author} {\bibfnamefont {S.~C.}\ \bibnamefont
  {Takatori}}, \bibinfo {author} {\bibfnamefont {W.}~\bibnamefont {Yan}},\ and\
  \bibinfo {author} {\bibfnamefont {J.~F.}\ \bibnamefont {Brady}},\ }\bibfield
  {title} {\bibinfo {title} {Swim pressure: Stress generation in active
  matter},\ }\href {https://doi.org/10.1103/PhysRevLett.113.028103} {\bibfield
  {journal} {\bibinfo  {journal} {Phys. Rev. Lett.}\ }\textbf {\bibinfo
  {volume} {113}},\ \bibinfo {pages} {028103} (\bibinfo {year}
  {2014})}\BibitemShut {NoStop}%
\bibitem [{\citenamefont {Nikola}\ \emph {et~al.}(2016)\citenamefont {Nikola},
  \citenamefont {Solon}, \citenamefont {Kafri}, \citenamefont {Kardar},
  \citenamefont {Tailleur},\ and\ \citenamefont {Voituriez}}]{Nikola2016aps}%
  \BibitemOpen
  \bibfield  {author} {\bibinfo {author} {\bibfnamefont {N.}~\bibnamefont
  {Nikola}}, \bibinfo {author} {\bibfnamefont {A.~P.}\ \bibnamefont {Solon}},
  \bibinfo {author} {\bibfnamefont {Y.}~\bibnamefont {Kafri}}, \bibinfo
  {author} {\bibfnamefont {M.}~\bibnamefont {Kardar}}, \bibinfo {author}
  {\bibfnamefont {J.}~\bibnamefont {Tailleur}},\ and\ \bibinfo {author}
  {\bibfnamefont {R.}~\bibnamefont {Voituriez}},\ }\bibfield  {title} {\bibinfo
  {title} {Active particles with soft and curved walls: Equation of state,
  ratchets, and instabilities},\ }\href
  {https://doi.org/10.1103/PhysRevLett.117.098001} {\bibfield  {journal}
  {\bibinfo  {journal} {Phys. Rev. Lett.}\ }\textbf {\bibinfo {volume} {117}},\
  \bibinfo {pages} {098001} (\bibinfo {year} {2016})}\BibitemShut {NoStop}%
\bibitem [{\citenamefont {Digregorio}\ \emph {et~al.}(2018)\citenamefont
  {Digregorio}, \citenamefont {Levis}, \citenamefont {Suma}, \citenamefont
  {Cugliandolo}, \citenamefont {Gonnella},\ and\ \citenamefont
  {Pagonabarraga}}]{Digregorio2018fpd}%
  \BibitemOpen
  \bibfield  {author} {\bibinfo {author} {\bibfnamefont {P.}~\bibnamefont
  {Digregorio}}, \bibinfo {author} {\bibfnamefont {D.}~\bibnamefont {Levis}},
  \bibinfo {author} {\bibfnamefont {A.}~\bibnamefont {Suma}}, \bibinfo {author}
  {\bibfnamefont {L.~F.}\ \bibnamefont {Cugliandolo}}, \bibinfo {author}
  {\bibfnamefont {G.}~\bibnamefont {Gonnella}},\ and\ \bibinfo {author}
  {\bibfnamefont {I.}~\bibnamefont {Pagonabarraga}},\ }\bibfield  {title}
  {\bibinfo {title} {Full phase diagram of active brownian disks: From melting
  to motility-induced phase separation},\ }\href
  {https://doi.org/10.1103/PhysRevLett.121.098003} {\bibfield  {journal}
  {\bibinfo  {journal} {Phys. Rev. Lett.}\ }\textbf {\bibinfo {volume} {121}},\
  \bibinfo {pages} {098003} (\bibinfo {year} {2018})}\BibitemShut {NoStop}%
\bibitem [{\citenamefont {Speck}(2016)}]{Speck2016sta}%
  \BibitemOpen
  \bibfield  {author} {\bibinfo {author} {\bibfnamefont {T.}~\bibnamefont
  {Speck}},\ }\bibfield  {title} {\bibinfo {title} {Stochastic thermodynamics
  for active matter},\ }\href {http://stacks.iop.org/0295-5075/114/i=3/a=30006}
  {\bibfield  {journal} {\bibinfo  {journal} {Europhys. Lett.}\ }\textbf
  {\bibinfo {volume} {114}},\ \bibinfo {pages} {30006} (\bibinfo {year}
  {2016})}\BibitemShut {NoStop}%
\bibitem [{\citenamefont {Speck}\ and\ \citenamefont
  {Jack}(2016)}]{Speck2016ibp}%
  \BibitemOpen
  \bibfield  {author} {\bibinfo {author} {\bibfnamefont {T.}~\bibnamefont
  {Speck}}\ and\ \bibinfo {author} {\bibfnamefont {R.~L.}\ \bibnamefont
  {Jack}},\ }\bibfield  {title} {\bibinfo {title} {Ideal bulk pressure of
  active brownian particles},\ }\href
  {https://doi.org/10.1103/PhysRevE.93.062605} {\bibfield  {journal} {\bibinfo
  {journal} {Phys. Rev. E}\ }\textbf {\bibinfo {volume} {93}},\ \bibinfo
  {pages} {062605} (\bibinfo {year} {2016})}\BibitemShut {NoStop}%
\bibitem [{\citenamefont {Levis}\ \emph {et~al.}(2017)\citenamefont {Levis},
  \citenamefont {Codina},\ and\ \citenamefont {Pagonabarraga}}]{Levis2017abe}%
  \BibitemOpen
  \bibfield  {author} {\bibinfo {author} {\bibfnamefont {D.}~\bibnamefont
  {Levis}}, \bibinfo {author} {\bibfnamefont {J.}~\bibnamefont {Codina}},\ and\
  \bibinfo {author} {\bibfnamefont {I.}~\bibnamefont {Pagonabarraga}},\
  }\bibfield  {title} {\bibinfo {title} {Active brownian equation of state:
  metastability and phase coexistence},\ }\href
  {https://doi.org/10.1039/C7SM01504F} {\bibfield  {journal} {\bibinfo
  {journal} {Soft Matter}\ }\textbf {\bibinfo {volume} {13}},\ \bibinfo {pages}
  {8113} (\bibinfo {year} {2017})}\BibitemShut {NoStop}%
\bibitem [{Not()}]{Note:ReferencePressure}%
  \BibitemOpen
  \href@noop {} {}\bibinfo {note} {We note that the irrelevant reference
  pressure from the surrounding fluid is deliberately ignored.}\BibitemShut
  {Stop}%
\bibitem [{\citenamefont {Yan}\ and\ \citenamefont {Brady}(2015)}]{Yan2015sfa}%
  \BibitemOpen
  \bibfield  {author} {\bibinfo {author} {\bibfnamefont {W.}~\bibnamefont
  {Yan}}\ and\ \bibinfo {author} {\bibfnamefont {J.~F.}\ \bibnamefont
  {Brady}},\ }\bibfield  {title} {\bibinfo {title} {The swim force as a body
  force},\ }\href {https://doi.org/10.1039/c5sm01318f} {\bibfield  {journal}
  {\bibinfo  {journal} {Soft Matter}\ }\textbf {\bibinfo {volume} {11}},\
  \bibinfo {pages} {6235} (\bibinfo {year} {2015})},\ \Eprint
  {https://arxiv.org/abs/https://pubs.rsc.org/sm/article-pdf/11/31/6235/4604604/c5sm01318f.pdf}
  {https://pubs.rsc.org/sm/article-pdf/11/31/6235/4604604/c5sm01318f.pdf}
  \BibitemShut {NoStop}%
\bibitem [{\citenamefont {Bialk\'e}\ \emph {et~al.}(2015)\citenamefont
  {Bialk\'e}, \citenamefont {Siebert}, \citenamefont {L\"owen},\ and\
  \citenamefont {Speck}}]{Bialke:2015nit}%
  \BibitemOpen
  \bibfield  {author} {\bibinfo {author} {\bibfnamefont {J.}~\bibnamefont
  {Bialk\'e}}, \bibinfo {author} {\bibfnamefont {J.~T.}\ \bibnamefont
  {Siebert}}, \bibinfo {author} {\bibfnamefont {H.}~\bibnamefont {L\"owen}},\
  and\ \bibinfo {author} {\bibfnamefont {T.}~\bibnamefont {Speck}},\ }\bibfield
   {title} {\bibinfo {title} {Negative interfacial tension in phase-separated
  active brownian particles},\ }\href
  {https://doi.org/10.1103/PhysRevLett.115.098301} {\bibfield  {journal}
  {\bibinfo  {journal} {Phys. Rev. Lett.}\ }\textbf {\bibinfo {volume} {115}},\
  \bibinfo {pages} {098301} (\bibinfo {year} {2015})}\BibitemShut {NoStop}%
\bibitem [{\citenamefont {Junot}\ \emph {et~al.}(2017)\citenamefont {Junot},
  \citenamefont {Briand}, \citenamefont {Ledesma-Alonso},\ and\ \citenamefont
  {Dauchot}}]{Junot2017avp}%
  \BibitemOpen
  \bibfield  {author} {\bibinfo {author} {\bibfnamefont {G.}~\bibnamefont
  {Junot}}, \bibinfo {author} {\bibfnamefont {G.}~\bibnamefont {Briand}},
  \bibinfo {author} {\bibfnamefont {R.}~\bibnamefont {Ledesma-Alonso}},\ and\
  \bibinfo {author} {\bibfnamefont {O.}~\bibnamefont {Dauchot}},\ }\bibfield
  {title} {\bibinfo {title} {Active versus passive hard disks against a
  membrane: Mechanical pressure and instability},\ }\href
  {https://doi.org/10.1103/PhysRevLett.119.028002} {\bibfield  {journal}
  {\bibinfo  {journal} {Phys. Rev. Lett.}\ }\textbf {\bibinfo {volume} {119}},\
  \bibinfo {pages} {028002} (\bibinfo {year} {2017})}\BibitemShut {NoStop}%
\bibitem [{\citenamefont {Speck}(2020)}]{Speck2020cfs}%
  \BibitemOpen
  \bibfield  {author} {\bibinfo {author} {\bibfnamefont {T.}~\bibnamefont
  {Speck}},\ }\bibfield  {title} {\bibinfo {title} {Collective forces in scalar
  active matter},\ }\href {https://doi.org/10.1039/d0sm00176g} {\bibfield
  {journal} {\bibinfo  {journal} {Soft Matter}\ }\textbf {\bibinfo {volume}
  {16}},\ \bibinfo {pages} {2652} (\bibinfo {year} {2020})},\ \Eprint
  {https://arxiv.org/abs/https://pubs.rsc.org/sm/article-pdf/16/11/2652/7565925/d0sm00176g.pdf}
  {https://pubs.rsc.org/sm/article-pdf/16/11/2652/7565925/d0sm00176g.pdf}
  \BibitemShut {NoStop}%
\bibitem [{\citenamefont {Sekimoto}(2010)}]{Sekimoto2010se}%
  \BibitemOpen
  \bibfield  {author} {\bibinfo {author} {\bibfnamefont {K.}~\bibnamefont
  {Sekimoto}},\ }\href@noop {} {\emph {\bibinfo {title} {Stochastic
  energetics}}},\ \bibinfo {series} {Lecture Notes in Physics}, Vol.\ \bibinfo
  {volume} {799}\ (\bibinfo  {publisher} {Springer},\ \bibinfo {year}
  {2010})\BibitemShut {NoStop}%
\bibitem [{\citenamefont {Seifert}(2012)}]{Seifert2012stf}%
  \BibitemOpen
  \bibfield  {author} {\bibinfo {author} {\bibfnamefont {U.}~\bibnamefont
  {Seifert}},\ }\bibfield  {title} {\bibinfo {title} {Stochastic
  thermodynamics, fluctuation theorems and molecular machines},\ }\href
  {http://stacks.iop.org/0034-4885/75/i=12/a=126001} {\bibfield  {journal}
  {\bibinfo  {journal} {Rep. Prog. Phys.}\ }\textbf {\bibinfo {volume} {75}},\
  \bibinfo {pages} {126001} (\bibinfo {year} {2012})}\BibitemShut {NoStop}%
\bibitem [{\citenamefont {Peliti}\ and\ \citenamefont
  {Pigolotti}(2021)}]{Peliti2021sti}%
  \BibitemOpen
  \bibfield  {author} {\bibinfo {author} {\bibfnamefont {L.}~\bibnamefont
  {Peliti}}\ and\ \bibinfo {author} {\bibfnamefont {S.}~\bibnamefont
  {Pigolotti}},\ }\href@noop {} {\emph {\bibinfo {title} {{Stochastic
  Thermodynamics: An Introduction}}}}\ (\bibinfo  {publisher} {Princeton
  University Press, Princeton, Oxford},\ \bibinfo {year} {2021})\BibitemShut
  {NoStop}%
\bibitem [{\citenamefont {Shiraishi}(2023)}]{Shiraishi2023ist}%
  \BibitemOpen
  \bibfield  {author} {\bibinfo {author} {\bibfnamefont {N.}~\bibnamefont
  {Shiraishi}},\ }\href
  {https://doi.org/https://doi.org/10.1007/978-981-19-8186-9} {\emph {\bibinfo
  {title} {{An Introduction to Stochastic Thermodynamics: From Basic to
  Advanced}}}},\ \bibinfo {edition} {1st}\ ed.,\ Fundamental Theories of
  Physics\ (\bibinfo  {publisher} {Springer Singapore},\ \bibinfo {year}
  {2023})\BibitemShut {NoStop}%
\bibitem [{\citenamefont {Jarzynski}(2011)}]{Jarzynski2011eii}%
  \BibitemOpen
  \bibfield  {author} {\bibinfo {author} {\bibfnamefont {C.}~\bibnamefont
  {Jarzynski}},\ }\bibfield  {title} {\bibinfo {title} {{E}qualities and
  {I}nequalities: {I}rreversibility and the {S}econd {L}aw of {T}hermodynamics
  at the {N}anoscale},\ }\href@noop {} {\bibfield  {journal} {\bibinfo
  {journal} {Annu. Rev. Condens. Matter Phys.}\ }\textbf {\bibinfo {volume}
  {2}},\ \bibinfo {pages} {7.1} (\bibinfo {year} {2011})}\BibitemShut {NoStop}%
\bibitem [{\citenamefont {Fily}\ and\ \citenamefont
  {Marchetti}(2012)}]{Fily2012aps}%
  \BibitemOpen
  \bibfield  {author} {\bibinfo {author} {\bibfnamefont {Y.}~\bibnamefont
  {Fily}}\ and\ \bibinfo {author} {\bibfnamefont {M.~C.}\ \bibnamefont
  {Marchetti}},\ }\bibfield  {title} {\bibinfo {title} {Athermal phase
  separation of self-propelled particles with no alignment},\ }\href
  {https://doi.org/10.1103/PhysRevLett.108.235702} {\bibfield  {journal}
  {\bibinfo  {journal} {Phys. Rev. Lett.}\ }\textbf {\bibinfo {volume} {108}},\
  \bibinfo {pages} {235702} (\bibinfo {year} {2012})}\BibitemShut {NoStop}%
\bibitem [{\citenamefont {Cates}\ and\ \citenamefont
  {Tailleur}(2013)}]{Cates2013waa}%
  \BibitemOpen
  \bibfield  {author} {\bibinfo {author} {\bibfnamefont {M.~E.}\ \bibnamefont
  {Cates}}\ and\ \bibinfo {author} {\bibfnamefont {J.}~\bibnamefont
  {Tailleur}},\ }\bibfield  {title} {\bibinfo {title} {When are active brownian
  particles and run-and-tumble particles equivalent? consequences for
  motility-induced phase separation},\ }\href
  {https://doi.org/10.1209/0295-5075/101/20010} {\bibfield  {journal} {\bibinfo
   {journal} {Europhysics Letters}\ }\textbf {\bibinfo {volume} {101}},\
  \bibinfo {pages} {20010} (\bibinfo {year} {2013})}\BibitemShut {NoStop}%
\bibitem [{\citenamefont {Farage}\ \emph {et~al.}(2015)\citenamefont {Farage},
  \citenamefont {Krinninger},\ and\ \citenamefont {Brader}}]{Farage2015eia}%
  \BibitemOpen
  \bibfield  {author} {\bibinfo {author} {\bibfnamefont {T.~F.~F.}\
  \bibnamefont {Farage}}, \bibinfo {author} {\bibfnamefont {P.}~\bibnamefont
  {Krinninger}},\ and\ \bibinfo {author} {\bibfnamefont {J.~M.}\ \bibnamefont
  {Brader}},\ }\bibfield  {title} {\bibinfo {title} {Effective interactions in
  active brownian suspensions},\ }\href
  {https://doi.org/10.1103/PhysRevE.91.042310} {\bibfield  {journal} {\bibinfo
  {journal} {Phys. Rev. E}\ }\textbf {\bibinfo {volume} {91}},\ \bibinfo
  {pages} {042310} (\bibinfo {year} {2015})}\BibitemShut {NoStop}%
\bibitem [{\citenamefont {Caprini}\ \emph {et~al.}(2019)\citenamefont
  {Caprini}, \citenamefont {Hernández-García}, \citenamefont {López},\ and\
  \citenamefont {Marini Bettolo~Marconi}}]{Caprini2019cst}%
  \BibitemOpen
  \bibfield  {author} {\bibinfo {author} {\bibfnamefont {L.}~\bibnamefont
  {Caprini}}, \bibinfo {author} {\bibfnamefont {E.}~\bibnamefont
  {Hernández-García}}, \bibinfo {author} {\bibfnamefont {C.}~\bibnamefont
  {López}},\ and\ \bibinfo {author} {\bibfnamefont {U.}~\bibnamefont {Marini
  Bettolo~Marconi}},\ }\bibfield  {title} {\bibinfo {title} {A comparative
  study between two models of active cluster crystals},\ }\href
  {https://www.ncbi.nlm.nih.gov/pmc/articles/PMC6853940/} {\bibfield  {journal}
  {\bibinfo  {journal} {Sci. Rep.}\ }\textbf {\bibinfo {volume} {9}},\ \bibinfo
  {pages} {16687} (\bibinfo {year} {2019})}\BibitemShut {NoStop}%
\bibitem [{\citenamefont {Krishnamurthy}\ \emph {et~al.}(2016)\citenamefont
  {Krishnamurthy}, \citenamefont {Ghosh}, \citenamefont {Chatterji},
  \citenamefont {Ganapathy},\ and\ \citenamefont
  {Sood}}]{Krishnamurthy2016msh}%
  \BibitemOpen
  \bibfield  {author} {\bibinfo {author} {\bibfnamefont {S.}~\bibnamefont
  {Krishnamurthy}}, \bibinfo {author} {\bibfnamefont {S.}~\bibnamefont
  {Ghosh}}, \bibinfo {author} {\bibfnamefont {D.}~\bibnamefont {Chatterji}},
  \bibinfo {author} {\bibfnamefont {R.}~\bibnamefont {Ganapathy}},\ and\
  \bibinfo {author} {\bibfnamefont {A.~K.}\ \bibnamefont {Sood}},\ }\bibfield
  {title} {\bibinfo {title} {A micrometre-sized heat engine operating between
  bacterial reservoirs},\ }\href {https://doi.org/10.1038/nphys3870} {\bibfield
   {journal} {\bibinfo  {journal} {Nat. Phys.}\ }\textbf {\bibinfo {volume}
  {12}},\ \bibinfo {pages} {1134} (\bibinfo {year} {2016})}\BibitemShut
  {NoStop}%
\bibitem [{\citenamefont {Martin}\ \emph {et~al.}(2018)\citenamefont {Martin},
  \citenamefont {Nardini}, \citenamefont {Cates},\ and\ \citenamefont
  {Fodor}}]{Martin2018emp}%
  \BibitemOpen
  \bibfield  {author} {\bibinfo {author} {\bibfnamefont {D.}~\bibnamefont
  {Martin}}, \bibinfo {author} {\bibfnamefont {C.}~\bibnamefont {Nardini}},
  \bibinfo {author} {\bibfnamefont {M.~E.}\ \bibnamefont {Cates}},\ and\
  \bibinfo {author} {\bibfnamefont {E.}~\bibnamefont {Fodor}},\ }\bibfield
  {title} {\bibinfo {title} {Extracting maximum power from active colloidal
  heat engines},\ }\href {http://stacks.iop.org/0295-5075/121/i=6/a=60005}
  {\bibfield  {journal} {\bibinfo  {journal} {EPL (Europhysics Letters)}\
  }\textbf {\bibinfo {volume} {121}},\ \bibinfo {pages} {60005} (\bibinfo
  {year} {2018})}\BibitemShut {NoStop}%
\bibitem [{\citenamefont {Pietzonka}\ \emph {et~al.}(2019)\citenamefont
  {Pietzonka}, \citenamefont {Fodor}, \citenamefont {Lohrmann}, \citenamefont
  {Cates},\ and\ \citenamefont {Seifert}}]{Pietzonka2019aed}%
  \BibitemOpen
  \bibfield  {author} {\bibinfo {author} {\bibfnamefont {P.}~\bibnamefont
  {Pietzonka}}, \bibinfo {author} {\bibfnamefont {E.}~\bibnamefont {Fodor}},
  \bibinfo {author} {\bibfnamefont {C.}~\bibnamefont {Lohrmann}}, \bibinfo
  {author} {\bibfnamefont {M.~E.}\ \bibnamefont {Cates}},\ and\ \bibinfo
  {author} {\bibfnamefont {U.}~\bibnamefont {Seifert}},\ }\bibfield  {title}
  {\bibinfo {title} {Autonomous engines driven by active matter: Energetics and
  design principles},\ }\href {https://doi.org/10.1103/PhysRevX.9.041032}
  {\bibfield  {journal} {\bibinfo  {journal} {Phys. Rev. X}\ }\textbf {\bibinfo
  {volume} {9}},\ \bibinfo {pages} {041032} (\bibinfo {year}
  {2019})}\BibitemShut {NoStop}%
\bibitem [{\citenamefont {Ekeh}\ \emph {et~al.}(2020)\citenamefont {Ekeh},
  \citenamefont {Cates},\ and\ \citenamefont {Fodor}}]{Ekeh2020tca}%
  \BibitemOpen
  \bibfield  {author} {\bibinfo {author} {\bibfnamefont {T.}~\bibnamefont
  {Ekeh}}, \bibinfo {author} {\bibfnamefont {M.~E.}\ \bibnamefont {Cates}},\
  and\ \bibinfo {author} {\bibfnamefont {E.}~\bibnamefont {Fodor}},\ }\bibfield
   {title} {\bibinfo {title} {Thermodynamic cycles with active matter},\ }\href
  {https://doi.org/10.1103/PhysRevE.102.010101} {\bibfield  {journal} {\bibinfo
   {journal} {Phys. Rev. E}\ }\textbf {\bibinfo {volume} {102}},\ \bibinfo
  {pages} {010101} (\bibinfo {year} {2020})}\BibitemShut {NoStop}%
\bibitem [{\citenamefont {Datta}\ \emph {et~al.}(2022)\citenamefont {Datta},
  \citenamefont {Pietzonka},\ and\ \citenamefont {Barato}}]{Datta2022sla}%
  \BibitemOpen
  \bibfield  {author} {\bibinfo {author} {\bibfnamefont {A.}~\bibnamefont
  {Datta}}, \bibinfo {author} {\bibfnamefont {P.}~\bibnamefont {Pietzonka}},\
  and\ \bibinfo {author} {\bibfnamefont {A.~C.}\ \bibnamefont {Barato}},\
  }\bibfield  {title} {\bibinfo {title} {Second law for active heat engines},\
  }\href {https://doi.org/10.1103/PhysRevX.12.031034} {\bibfield  {journal}
  {\bibinfo  {journal} {Phys. Rev. X}\ }\textbf {\bibinfo {volume} {12}},\
  \bibinfo {pages} {031034} (\bibinfo {year} {2022})}\BibitemShut {NoStop}%
\bibitem [{\citenamefont {Davis}\ \emph {et~al.}(2024)\citenamefont {Davis},
  \citenamefont {Proesmans},\ and\ \citenamefont {Fodor}}]{Davis2024amc}%
  \BibitemOpen
  \bibfield  {author} {\bibinfo {author} {\bibfnamefont {L.~K.}\ \bibnamefont
  {Davis}}, \bibinfo {author} {\bibfnamefont {K.}~\bibnamefont {Proesmans}},\
  and\ \bibinfo {author} {\bibfnamefont {E.}~\bibnamefont {Fodor}},\ }\bibfield
   {title} {\bibinfo {title} {Active matter under control: Insights from
  response theory},\ }\href {https://doi.org/10.1103/PhysRevX.14.011012}
  {\bibfield  {journal} {\bibinfo  {journal} {Phys. Rev. X}\ }\textbf {\bibinfo
  {volume} {14}},\ \bibinfo {pages} {011012} (\bibinfo {year}
  {2024})}\BibitemShut {NoStop}%
\bibitem [{\citenamefont {Garcia-Millan}\ \emph {et~al.}(2025)\citenamefont
  {Garcia-Millan}, \citenamefont {Sch\"uttler}, \citenamefont {Cates},\ and\
  \citenamefont {Loos}}]{GarciaMillan2025ocl}%
  \BibitemOpen
  \bibfield  {author} {\bibinfo {author} {\bibfnamefont {R.}~\bibnamefont
  {Garcia-Millan}}, \bibinfo {author} {\bibfnamefont {J.}~\bibnamefont
  {Sch\"uttler}}, \bibinfo {author} {\bibfnamefont {M.~E.}\ \bibnamefont
  {Cates}},\ and\ \bibinfo {author} {\bibfnamefont {S.~A.~M.}\ \bibnamefont
  {Loos}},\ }\bibfield  {title} {\bibinfo {title} {Optimal closed-loop control
  of active particles and a minimal information engine},\ }\href
  {https://doi.org/10.1103/fbgp-qpvv} {\bibfield  {journal} {\bibinfo
  {journal} {Phys. Rev. Lett.}\ }\textbf {\bibinfo {volume} {135}},\ \bibinfo
  {pages} {088301} (\bibinfo {year} {2025})}\BibitemShut {NoStop}%
\bibitem [{\citenamefont {Davis}(2026)}]{Davis2026omp}%
  \BibitemOpen
  \bibfield  {author} {\bibinfo {author} {\bibfnamefont {L.~K.}\ \bibnamefont
  {Davis}},\ }\href {https://arxiv.org/abs/2603.16778} {\bibinfo {title}
  {Optimal multi-parameter control of trapped active matter}} (\bibinfo {year}
  {2026}),\ \Eprint {https://arxiv.org/abs/2603.16778} {arXiv:2603.16778
  [cond-mat.soft]} \BibitemShut {NoStop}%
\bibitem [{\citenamefont {Gaspard}\ and\ \citenamefont
  {Kapral}(2018)}]{Gaspard2018fcs}%
  \BibitemOpen
  \bibfield  {author} {\bibinfo {author} {\bibfnamefont {P.}~\bibnamefont
  {Gaspard}}\ and\ \bibinfo {author} {\bibfnamefont {R.}~\bibnamefont
  {Kapral}},\ }\bibfield  {title} {\bibinfo {title} {Fluctuating
  chemohydrodynamics and the stochastic motion of self-diffusiophoretic
  particles},\ }\href {https://doi.org/10.1063/1.5020442} {\bibfield  {journal}
  {\bibinfo  {journal} {J. Chem. Phys.}\ }\textbf {\bibinfo {volume} {148}},\
  \bibinfo {pages} {134104} (\bibinfo {year} {2018})},\ \Eprint
  {https://arxiv.org/abs/https://pubs.aip.org/aip/jcp/article-pdf/doi/10.1063/1.5020442/15539084/134104\textunderscore
  1\textunderscore online.pdf}
  {https://pubs.aip.org/aip/jcp/article-pdf/doi/10.1063/1.5020442/15539084/134104\textunderscore
  1\textunderscore online.pdf} \BibitemShut {NoStop}%
\bibitem [{\citenamefont {Pietzonka}\ and\ \citenamefont
  {Seifert}(2018)}]{Pietzonka2018epa}%
  \BibitemOpen
  \bibfield  {author} {\bibinfo {author} {\bibfnamefont {P.}~\bibnamefont
  {Pietzonka}}\ and\ \bibinfo {author} {\bibfnamefont {U.}~\bibnamefont
  {Seifert}},\ }\bibfield  {title} {\bibinfo {title} {Entropy production of
  active particles and for particles in active baths},\ }\href
  {http://stacks.iop.org/1751-8121/51/i=1/a=01LT01} {\bibfield  {journal}
  {\bibinfo  {journal} {J. Phys. A: Math. Theor.}\ }\textbf {\bibinfo {volume}
  {51}},\ \bibinfo {pages} {01LT01} (\bibinfo {year} {2018})}\BibitemShut
  {NoStop}%
\bibitem [{\citenamefont {Gaspard}\ and\ \citenamefont
  {Kapral}(2019)}]{Gaspard2019tsm}%
  \BibitemOpen
  \bibfield  {author} {\bibinfo {author} {\bibfnamefont {P.}~\bibnamefont
  {Gaspard}}\ and\ \bibinfo {author} {\bibfnamefont {R.}~\bibnamefont
  {Kapral}},\ }\bibfield  {title} {\bibinfo {title} {Thermodynamics and
  statistical mechanics of chemically powered synthetic nanomotors},\ }\href
  {https://doi.org/10.1080/23746149.2019.1602480} {\bibfield  {journal}
  {\bibinfo  {journal} {Advances in Physics: X}\ }\textbf {\bibinfo {volume}
  {4}},\ \bibinfo {pages} {1602480} (\bibinfo {year} {2019})},\ \Eprint
  {https://arxiv.org/abs/https://doi.org/10.1080/23746149.2019.1602480}
  {https://doi.org/10.1080/23746149.2019.1602480} \BibitemShut {NoStop}%
\bibitem [{\citenamefont {Fritz}\ and\ \citenamefont
  {Seifert}(2023)}]{Fritz2023tcm}%
  \BibitemOpen
  \bibfield  {author} {\bibinfo {author} {\bibfnamefont {J.~H.}\ \bibnamefont
  {Fritz}}\ and\ \bibinfo {author} {\bibfnamefont {U.}~\bibnamefont
  {Seifert}},\ }\bibfield  {title} {\bibinfo {title} {Thermodynamically
  consistent model of an active ornstein–uhlenbeck particle},\ }\href
  {https://doi.org/10.1088/1742-5468/acf70c} {\bibfield  {journal} {\bibinfo
  {journal} {J. Stat. Mech: Theory Exp.}\ }\textbf {\bibinfo {volume} {2023}},\
  \bibinfo {pages} {093204} (\bibinfo {year} {2023})}\BibitemShut {NoStop}%
\bibitem [{\citenamefont {Chatzittofi}\ \emph {et~al.}(2024)\citenamefont
  {Chatzittofi}, \citenamefont {Agudo-Canalejo},\ and\ \citenamefont
  {Golestanian}}]{Chatzittofi2024ept}%
  \BibitemOpen
  \bibfield  {author} {\bibinfo {author} {\bibfnamefont {M.}~\bibnamefont
  {Chatzittofi}}, \bibinfo {author} {\bibfnamefont {J.}~\bibnamefont
  {Agudo-Canalejo}},\ and\ \bibinfo {author} {\bibfnamefont {R.}~\bibnamefont
  {Golestanian}},\ }\bibfield  {title} {\bibinfo {title} {Entropy production
  and thermodynamic inference for stochastic microswimmers},\ }\href
  {https://doi.org/10.1103/PhysRevResearch.6.L022044} {\bibfield  {journal}
  {\bibinfo  {journal} {Phys. Rev. Res.}\ }\textbf {\bibinfo {volume} {6}},\
  \bibinfo {pages} {L022044} (\bibinfo {year} {2024})}\BibitemShut {NoStop}%
\bibitem [{\citenamefont {King}\ \emph {et~al.}(2017)\citenamefont {King},
  \citenamefont {Butcher},\ and\ \citenamefont {Zalewski}}]{King2017ahp}%
  \BibitemOpen
  \bibfield  {author} {\bibinfo {author} {\bibfnamefont {T.}~\bibnamefont
  {King}}, \bibinfo {author} {\bibfnamefont {S.}~\bibnamefont {Butcher}},\ and\
  \bibinfo {author} {\bibfnamefont {L.}~\bibnamefont {Zalewski}},\ }\href
  {https://doi.org/10.5281/zenodo.438045} {\emph {\bibinfo {title} {Apocrita -
  High Performance Computing Cluster for Queen Mary University of London}}}
  (\bibinfo {year} {2017})\BibitemShut {NoStop}%
\bibitem [{\citenamefont {El-Mikkawy}(2004)}]{ELMIKKAWY2004669}%
  \BibitemOpen
  \bibfield  {author} {\bibinfo {author} {\bibfnamefont {M.~E.}\ \bibnamefont
  {El-Mikkawy}},\ }\bibfield  {title} {\bibinfo {title} {On the inverse of a
  general tridiagonal matrix},\ }\href
  {https://doi.org/https://doi.org/10.1016/S0096-3003(03)00298-4} {\bibfield
  {journal} {\bibinfo  {journal} {Applied Mathematics and Computation}\
  }\textbf {\bibinfo {volume} {150}},\ \bibinfo {pages} {669} (\bibinfo {year}
  {2004})}\BibitemShut {NoStop}%
\bibitem [{\citenamefont {Shukla}\ and\ \citenamefont
  {Swaminathan}(2023)}]{Shukla2023}%
  \BibitemOpen
  \bibfield  {author} {\bibinfo {author} {\bibfnamefont {V.}~\bibnamefont
  {Shukla}}\ and\ \bibinfo {author} {\bibfnamefont {A.}~\bibnamefont
  {Swaminathan}},\ }\bibfield  {title} {\bibinfo {title} {Spectral
  transformation associated with a perturbed {\$}{\$}r{\_}i{\$}{\$}type
  recurrence relation},\ }\href {https://doi.org/10.1007/s40840-023-01561-8}
  {\bibfield  {journal} {\bibinfo  {journal} {Bulletin of the Malaysian
  Mathematical Sciences Society}\ }\textbf {\bibinfo {volume} {46}},\ \bibinfo
  {pages} {169} (\bibinfo {year} {2023})}\BibitemShut {NoStop}%
\bibitem [{\citenamefont {Coussement}\ \emph {et~al.}(2002)\citenamefont
  {Coussement}, \citenamefont {Kuijlaars},\ and\ \citenamefont
  {Van~Assche}}]{Coussement2002}%
  \BibitemOpen
  \bibfield  {author} {\bibinfo {author} {\bibfnamefont {J.}~\bibnamefont
  {Coussement}}, \bibinfo {author} {\bibfnamefont {A.}~\bibnamefont
  {Kuijlaars}},\ and\ \bibinfo {author} {\bibfnamefont {W.}~\bibnamefont
  {Van~Assche}},\ }\bibfield  {title} {\bibinfo {title} {Direct and inverse
  spectral transform for the relativistic toda lattice and the connection with
  laurent orthogonal polynomials},\ }\href
  {https://doi.org/10.1088/0266-5611/18/3/325} {\bibfield  {journal} {\bibinfo
  {journal} {Inverse Problems}\ }\textbf {\bibinfo {volume} {18}} (\bibinfo
  {year} {2002})}\BibitemShut {NoStop}%
\bibitem [{\citenamefont {Behringer}\ and\ \citenamefont
  {Eichhorn}(2012)}]{Behringer2012bds}%
  \BibitemOpen
  \bibfield  {author} {\bibinfo {author} {\bibfnamefont {H.}~\bibnamefont
  {Behringer}}\ and\ \bibinfo {author} {\bibfnamefont {R.}~\bibnamefont
  {Eichhorn}},\ }\bibfield  {title} {\bibinfo {title} {{Brownian dynamics
  simulations with hard-body interactions: Spherical particles}},\ }\href
  {https://doi.org/10.1063/1.4761827} {\bibfield  {journal} {\bibinfo
  {journal} {J. Chem. Phys.}\ }\textbf {\bibinfo {volume} {137}},\ \bibinfo
  {pages} {164108} (\bibinfo {year} {2012})},\ \Eprint
  {https://arxiv.org/abs/https://doi.org/10.1063/1.4761827}
  {https://doi.org/10.1063/1.4761827} \BibitemShut {NoStop}%
\end{thebibliography}

%

\clearpage

\onecolumngrid
\begin{center}
\ \\[10pt]
{\large\textbf{SUPPLEMENTAL MATERIAL}}
\end{center}
\vspace{25pt}
\twocolumngrid

\setcounter{section}{0}
\renewcommand{\thesection}{S\arabic{section}}
\setcounter{figure}{0}
\renewcommand{\thefigure}{S\arabic{figure}}
\setcounter{equation}{0}
\renewcommand{\theequation}{S\arabic{equation}}
\setcounter{table}{0}
\renewcommand{\thetable}{S\arabic{table}}

Labels of equations
and references in these Supplementary Notes are prefixed by a capital letter ``S'' (e.g., Fig.~S1, Eq.~(S3)).
Any plain labels (e.g., Fig.~1, Eq.~(3)) refer to the corresponding items in the main text.

\section{Steady-state solution for AOUPs}
\label{app:HermiteFunctionExpansion}

Here
we explicitly construct the steady-state probability densities for AOUPs in a box,
i.e., the solution of the system~\meqref{eq:ODE:X} of ordinary differential equations (ODEs)
with boundary conditions~\meqref{eq:BC:x}.
We work in dimensionless units throughout this section (without the explicit tilde notation)
and write $\alpha \equiv \Pe$ for clarity.

The Hermite function expansion~\meqref{eq:rho:ansatz} for the steady-state probability density $\rho(x, \eta)$ leads to the infinite ODE system~\meqref{eq:ODE:X} for the unknown functions $X_n(x)$ ($n = 0, 1, \ldots$).
Truncating at order $M$ by setting $X_n \equiv 0$ for $n > M$,
we can write this ODE system in matrix form as
\begin{equation}
\label{eq:ODE:X:mat}
    \bm X'' - \alpha \bm A \bm X' -  \bm D \bm X = 0 \,,
\end{equation}
where $\bm X := (X_0 \, \cdots \, X_M)^\T$, and $\bm A = (A_{mn})$ and $\bm D = (D_{mn})$ are $(M+1)\times(M+1)$ matrices defined by
\begin{align}
    A_{mn} &:= \delta_{m,n+1} + n \, \delta_{m, n-1} \,, \\
    D_{mn} &:= n \, \delta_{m, n}
\end{align}
for $m, n \in \{ 0, \ldots, M \}$.
Likewise,
the boundary conditions~\meqref{eq:BC:x} can be written as
\begin{equation}
\label{eq:BC:X:mat}
    \bm X'(x) - \alpha \bm A \bm X(x) = 0
    \quad \text{if} \quad
    x = 0, L \,.
\end{equation}
Since the ODE system~\eqref{eq:ODE:X:mat} is linear with constant coefficients,
we look for solutions of the form
\begin{equation}
\label{eq:ansatz:X}
    \bm X(x) = \bm C \, \e^{\lambda x}
\end{equation}
with a constant vector $\bm C = (C_0 \, \ldots \, C_M)^\T$.
Substituting this ansatz into~\eqref{eq:ODE:X:mat},
we obtain the linear system
\begin{equation}
\label{eq:lambda}
    (\lambda^2 \bm I - \lambda \alpha \bm A - \bm D) \, \bm C = 0 \,,
\end{equation}
where $\bm I$ denotes the identity matrix.
Since we are interested in values of $\lambda$ that allow for nontrivial solutions of this system,
we require
\begin{equation}
\label{eq:lambda:det}
    F_M(\mu = \lambda^2) := \det(\lambda^2 \bm I - \lambda \alpha \bm A - \bm D) = 0 \,.
\end{equation}
Thanks to the tridiagonal structure of $\lambda^2 \bm I - \lambda \alpha \bm A - \bm D$,
the determinants $F_M(\mu)$ satisfy a recurrence relation in $M$ \cite{ELMIKKAWY2004669},
namely
\begin{equation}
\label{eq:lambda:recurrence}
    F_M(\mu) = (\mu - M) \, F_{M-1}(\mu) - M \mu \alpha^2 F_{M-2}(\mu)
\end{equation}
with initial conditions $F_{0} = \mu$ and $F_{-1} = 1$ in the present case.
Consequently, $F_M(\mu)$ is a polynomial of order $M+1$ in $\mu = \lambda^2$.
One simple root of $F_M(\mu)$ is $\mu = 0$ (for $M \geq 0$).
The remaining $M$ roots are all positive and distinct \cite{Shukla2023, Coussement2002},
and we denote them by $\mu_1, \ldots, \mu_M$.
Eq.~\eqref{eq:lambda:det} thus has $2M + 1$ distinct solutions $\lambda^{(-M)}, \ldots, \lambda^{(M)}$,
namely $\lambda_0 := 0$ and $\lambda_{\pm m} := \pm \sqrt{\mu_m}$ ($m = 1, \ldots, M$).

For $\lambda_0 = 0$,
the corresponding solution $\bm C^{(0)}$ of~\eqref{eq:lambda} is
\begin{equation}
\label{eq:C0}
    \bm C^{(0)} = (1 \, 0 \, \cdots \, 0)^\T
\end{equation}
up to an arbitrary constant scaling factor.
For all other $m \neq 0$,
we can compute the coefficients of $\bm C^{(m)}$ recursively from~\eqref{eq:lambda},
\begin{subequations}
\label{eq:Ck}
\begin{align}
    C^{(m)}_0 &:= 1 \,, \\
\label{eq:Ck1}
    C^{(m)}_1 &:= \frac{\lambda_m}{\alpha} \,, \\
    C^{(m)}_{n+1} &:= \frac{(\lambda_m^2 - n) C^{(m)}_n - \lambda_m \alpha C^{(m)}_{n-1}}{ (n+1) \lambda_m \alpha} \,.
\end{align}
\end{subequations}
The symmetry $\lambda_{-m} = -\lambda_m$ furthermore implies
\begin{equation}
\label{eq:Ck:symmetry}
    C^{(-m)}_n = (-1)^n \, C^{(m)}_n \,.
\end{equation}
 
In summary, the general solution of Eq.~\eqref{eq:ODE:X:mat} thus takes the form
\begin{equation}
\begin{aligned}
\label{eq:X:general}
    \bm X(x) &= \sum_{m = -M}^M B_m \bm C^{(m)} \e^{\lambda_m x} \\
    &= B_0 \bm C^{(0)} + \sum_{m=1}^M \left[ B_m \bm C^{(m)} \e^{\lambda_m x} + B_{-m} \bm C^{(-m)} \e^{-\lambda_m x} \right] ,
\end{aligned}
\end{equation}
where $\lambda_m$ are the $2M + 1$ simple, pairwise distinct roots of $F_M(\lambda^2) = 0$, cf.\ Eqs.~\eqref{eq:lambda:det} and~\eqref{eq:lambda:recurrence},
and the $\bm C^{(m)}$ are constructed according to Eqs.~\eqref{eq:C0} and~\eqref{eq:Ck}.

The constants $B_m$ are determined by the boundary conditions~\eqref{eq:BC:X:mat}
and the normalization condition
\begin{equation}
\label{eq:norm:X}
    \int_0^L \d x \int_{-\infty}^\infty \d\eta \, \rho(x, \eta) = \int_0^L \d x \; X_0(x) = 1 \,,
\end{equation}
where the first equality follows after plugging in the ansatz~\meqref{eq:rho:ansatz} and exploiting the orthogonality property~\meqref{eq:He:Orthogonality} of the Hermite polynomials.
Substituting the general solution~\eqref{eq:X:general} into the boundary condition~\eqref{eq:BC:X:mat}
and exploiting that $(\lambda_m\bm I - \alpha \bm A) \bm C^{(m)} = \lambda_m^{-1} \bm D \bm C^{(m)}$ for $m \neq 0$ thanks to~\eqref{eq:lambda},
we obtain
\begin{equation}
\label{eq:BC:X:subs}
\begin{aligned}
    &-\alpha \bm A \bm C^{(0)} B_0 \\
    & \; + \sum_{m=1}^M \frac{\bm D}{\lambda_m} \left[ \bm C^{(m)} B_m \e^{\lambda_m x} - \bm C^{(-m)} B_{-m} \e^{-\lambda_m x} \right] = 0
\end{aligned}
\end{equation}
if $x = 0, L$.
Since the vectors $\bm C^{(m)}$ have $M+1$ components $C^{(m)}_n$,
the choices $x = 0$ and $x = L$ yield a total of $2(M + 1)$ equations.
However, the equations for the component index $n = 0$ are automatically satisfied, regardless of the choice of the $B_m$, because $D_{0j} = 0$ and $(\bm A \bm C^{(0)})_0 = 0$,
leaving only $2M$ nontrivial equations from the boundary conditions.
Together with the single equation~\eqref{eq:norm:X} from the normalization condition,
we thus have a total of $2M + 1$ linear equations to fix the $2M + 1$ coefficients $B_m$.
Writing the nontrivial equations~\eqref{eq:BC:X:subs} out in components for $x = 0$ and $x = L$
and using the symmetry~\eqref{eq:Ck:symmetry},
we find
\begin{align}
\label{eq:BC:X:subs:0}
    & \sum_{m=1}^M \frac{n}{\lambda_m} C^{(m)}_n \left[ B_m - (-1)^n B_{-m} \right] - \alpha \delta_{n, 1} B_0 = 0 \,, \\
\label{eq:BC:X:subs:L}
    & \sum_{m=1}^M \frac{n}{\lambda_m} C^{(m)}_n \left[ B_m \e^{\lambda_m L} - (-1)^n B_{-m} \e^{-\lambda_m L} \right] - \alpha \delta_{n,1} B_0 = 0 
\end{align}
with $n = 1, \ldots, M$.
For fixed $n$,
we can obtain two new, equivalent sets of equations by taking, respectively, the sum and difference of~\eqref{eq:BC:X:subs:0} and~\eqref{eq:BC:X:subs:L}.
The choice
\begin{equation}
\label{eq:BkSymmetry}
    B_{-m} = \e^{\lambda_m L} B_m
\end{equation}
for $m > 0$
then
satisfies the
sum equations for even $n$ 
and the difference equations for odd $n$,
effectively halving the number of equations (and variables).
Hence, there remain $M$ equations,
namely
\begin{align}
\label{eq:BC:X:subs:+}
     & \sum_{m=1}^M \frac{n}{\lambda_m} C^{(m)}_n B_m (1 + \e^{\lambda_m L}) - \alpha \delta_{n, 1} B_0 = 0 \quad (n \; \text{odd}) \,, \\
\label{eq:BC:X:subs:-}
     & \sum_{m=1}^M \frac{n}{\lambda_m} C^{(m)}_n B_m (1 - \e^{\lambda_m L}) = 0 \quad (n \; \text{even}) \,.
\end{align}
The normalization condition~\eqref{eq:norm:X}, in turn,
yields
\begin{equation}
\label{eq:norm:X:subs}
    B_0 L + 2 \sum_{m=1}^M \frac{1}{\lambda_m} B_m \left( \e^{\lambda_m L} - 1 \right) = 1 \,.
\end{equation}
The remaining system~\eqref{eq:BC:X:subs:+}-\eqref{eq:norm:X:subs} of $M + 1$ linear equations for the $M+1$ variables $B_0, \ldots, B_M$ must be solved numerically.

For instance, at truncation order $M = 1$,
solving~\eqref{eq:lambda:det} yields $\lambda_1 = \sqrt{1 + \alpha^2}$ [and $\lambda_0 = 0$, see below Eq.~\eqref{eq:lambda:recurrence}].
The coefficients $B_0$ and $B_1$ satisfy~\eqref{eq:BC:X:subs:+} and~\eqref{eq:norm:X:subs},
\begin{align}
    \alpha^2 B_0 - (1 + \e^{\lambda_1 L}) B_1 &= 0 \\
    L B_0 + \frac{2}{\lambda_1} \left( \e^{\lambda_k L} - 1 \right) B_1 &= 1 \,.
\end{align}
The solution is
\begin{align}
    B_0 &= \frac{1}{L} \left[ 1 + \alpha^2 \frac{\tanh(\lambda_1 L / 2)}{\lambda_1 L / 2} \right]^{-1} \,, \\
    B_1 &= \frac{\alpha^2}{1 + \e^{\lambda_1 L}} B_0 \,.
\end{align}

\section{Exact solution for RTPs in $d = 1$}
\label{app:ExactSolutionRTP}

Here we verify that Eq.~\meqref{eq:rho:x:M1} is the exact steady-state probability density of a run-and-tumple particle (RTP) on the interval $[0, L]$ with no-flux boundary conditions.
We also demonstrate that the formulation~\meqref{eq:IdealActiveGasLaw} of the ideal active gas law is exact in this case.
We work with dimensionless quantities throughout this section (without the explicit tilde notation) and denote $\alpha \equiv \Pe$ ($= v_0$ in dimensionless quantities) for clarity.

In $d = 1$,
RTPs can be modeled as two species, left- and right-moving particles,
which transform into each other at a rate $\ta^{-1}$.
In the steady state,
the probability densities $\rho_-(x)$ and $\rho_+(x)$ of the left- and right-moving particles, respectively,
satisfy the coupled Fokker-Planck equations \cite{GarciaMillan2021rtm}
\begin{equation}
\label{eq:RTP:FP}
    \rho_\pm'' \mp \alpha \rho_\pm' \mp \frac{1}{2} (\rho_+ - \rho_-) = 0 \,.
\end{equation}
Adding the ``$+$'' and ``$-$'' equations together and integrating once,
we obtain
\begin{equation}
\label{eq:RTP:LRsum}
    \rho' + \alpha (\rho_- - \rho_+) = C \,,
\end{equation}
where $C$ is a constant and
\begin{equation}
\label{eq:RTP:rho}
    \rho = \rho_+ + \rho_-
\end{equation}
is the total position density.
On the left-hand side of Eq.~\eqref{eq:RTP:LRsum}, we recognize the total probability current $J = J_+ + J_-$
with $J_\pm = (\mp \alpha - \frac{\d}{\d x}) \rho_\pm$.
The no-flux boundary conditions thus imply $C = 0$
and therefore
\begin{equation}
\label{eq:RTP:diff}
    \rho_+ - \rho_- = \frac{\rho'}{\alpha} \,.
\end{equation}
Combining~\eqref{eq:RTP:rho} and~\eqref{eq:RTP:diff},
we conclude that
\begin{equation}
    \rho_\pm = \frac{1}{2} \left( \rho \pm \frac{\rho'}{\alpha} \right) .
\end{equation}
Upon substitution of Eq.~\meqref{eq:rho:x:M1} for $\rho$,
we find
\begin{equation}
\label{eq:RTP:rhopm}
    \rho_\pm(x) = \frac{\cosh\ell + \alpha^2 \cosh(\beta [x-\frac{L}{2}]) \pm \alpha \beta \sinh( \beta [x-\frac{L}{2}] )}{2 L (1 + \alpha^2 \vartheta) \cosh\ell}
\end{equation}
with $\vartheta$ and $\ell$ as defined below Eq.~\meqref{eq:IdealActiveGasLaw}, $\vartheta := \frac{\tanh\ell}{\ell}$ and $\ell := L \sqrt{\frac{1 + \Pe^2}{4 D \ta}}$, and $\beta := \sqrt{1 + \alpha^2}$.
A straightforward calculation confirms that this is indeed a solution of~\eqref{eq:RTP:FP},
i.e., $\rho(x)$ from~\meqref{eq:rho:x:M1} is the total RTP probability density.

To verify the ideal active gas law~\meqref{eq:IdealActiveGasLaw},
we also need the correlation
\begin{equation}
\label{eq:RTP:xv}
    \< x v \> = \frac{1}{2} \left( \< x \,|\, v = + v_0 \> - \< x \,|\, v = -v_0 \> \right).
\end{equation}
The conditional probability densities are $\rho(x | v = \pm v_0) = 2 \rho_\pm(x)$.
Adopting~\eqref{eq:RTP:rhopm} to evaluate~\eqref{eq:RTP:xv},
we obtain
\begin{equation}
    \< x v \> = \frac{\alpha^2 (1 - \vartheta)}{1 + \alpha^2 \vartheta} \,.
\end{equation}
Note that this is the result for dimensionless variables;
to cast it to dimensionful quantities, we need to multiply it by the diffusion coefficient $D$
[see above Eq.~\meqref{eq:xeta}].
Substituting the so-rescaled $\< x v \>$ into~\meqref{eq:pressure},
we recover the ideal active gas law~\meqref{eq:IdealActiveGasLaw}.

\section{Higher-order approximations of the ideal active gas law for AOUPs}
\label{app:HigherOrderAOUPLaw}

We analyze the structure of the ideal active gas law~\meqref{eq:IdealActiveGasLaw} for AOUPs using higher-order truncations of the steady-state solution~\meqref{eq:rho:ansatz} in more detail.
The pertinent generalization to higher dimensions with different linear box sizes $L_k$ is straightforwardly obtained by carrying out the steps detailed in the following for each dimension separately,
resulting in pressure contributions $P_k$ for each dimension as in~\meqref{eq:Pk} and an ideal active gas law for the total pressure similar to~\meqref{eq:IdealActiveGasLaw:multi} by averaging over those contributions.

Working with dimensionless quantities,
we therefore revisit the correlation $\<x \eta\>$ from Eq.~\meqref{eq:xeta}.
Truncating the steady-state AOUP solution at order $M$,
we have [cf.\ Eqs.~\eqref{eq:Ck1}, \eqref{eq:X:general}, and~\eqref{eq:BkSymmetry} in Appendix~\ref{app:HermiteFunctionExpansion}]
\begin{equation}
    X_1(x)
        = \frac{1}{\alpha} \sum_{m=1}^M B_m \lambda_m \left[ \e^{\lambda_m x} - \e^{\lambda_m (L-x)} \right] ,
\end{equation}
where $\alpha \equiv \Pe$ as before and the $B_m$ are obtained by solving the linear system of Eqs.~\eqref{eq:BC:X:subs:+}--\eqref{eq:norm:X:subs}.
Substituting into~\meqref{eq:xeta} and performing the $x$ integral,
we obtain
\begin{equation}
\label{eq:xeta:HigherOrder}
    \< x \eta \>
        = \frac{L}{\alpha} \sum_{m=1}^M B_m \left( 1 + \e^{2 \ell^{(m)}} \right) \left( 1 - \frac{\tanh \ell^{(m)}}{\ell^{(m)}} \right) ,
\end{equation}
where $\ell^{(m)} := \lambda_m L / 2$.
Defining
\begin{equation}
\label{eq:wk}
    w_n := \frac{1 + \e^{2 \ell^{(n)}}}{\alpha^2 B_0} B_n
\end{equation}
and
\begin{equation}
    H_{mn} := \frac{m \alpha}{\lambda_n} C^{(m)}_n
    \times
    \begin{cases}
        1 \,, & m \text{ odd} \,, \\
        \tanh \ell^{(n)} \,, & m \text{ even} \,,
    \end{cases}
\end{equation}
for $m, n = 1, \ldots, M$,
Eqs.~\eqref{eq:BC:X:subs:+} and~\eqref{eq:BC:X:subs:-}
can be written as
\begin{equation}
\label{eq:wk:system}
    \bm H \bm w = (1 \; 0 \; \cdots \; 0)^\T \,,
\end{equation}
where $\bm H = (H_{mn})$ and $\bm w = (w_n)$.
Recalling Eq.~\eqref{eq:Ck1},
we observe that $H_{1n} = 1$ for all $n$,
hence
\begin{equation}
\label{eq:wk:norm}
    (\bm H \bm w)_1 = \sum_{n=1}^M w_n = 1
\end{equation}
as stated below~\meqref{eq:IdealActiveGasLaw:HigherOrderTheta}.
By solving the full system~\eqref{eq:wk:system},
the weights $w_n$ can be calculated explicitly.
The remaining coefficient $B_0$ follows from the normalization condition~\eqref{eq:norm:X},
\begin{equation}
    B_0 = \frac{1}{L} \left[ 1 + \alpha^2 \sum_{m=1}^M w_m \frac{\tanh \ell^{(m)}}{\ell^{(m)}} \right]^{-1}
        = \frac{1}{L (1 + \alpha^2 \vartheta)}
\end{equation}
with 
\begin{equation}
\label{eq:IdealActiveGasLaw:HigherOrderTheta:app}
    \vartheta := \sum_{m=1}^M w_m \frac{\tanh \ell^{(m)}}{\ell^{(m)}} \,,
\end{equation}
cf.\ Eq.~\meqref{eq:IdealActiveGasLaw:HigherOrderTheta}.
Adopting Eqs.~\eqref{eq:wk}, \eqref{eq:wk:norm}, and~\eqref{eq:IdealActiveGasLaw:HigherOrderTheta:app},
we find from Eq.~\eqref{eq:xeta:HigherOrder} that
\begin{equation}
    \< x \eta \> = \frac{\alpha (1 - \vartheta)}{1 + \alpha^2 \vartheta} \,.
\end{equation}
Switching to dimensionful quantities and recalling $\< \bm x \cdot \bm v \> = d D \alpha \< \tilde x \tilde \eta \>$,
we then recover the form~\meqref{eq:IdealActiveGasLaw} for the ideal active gas law,
but with the higher-order approximation~\eqref{eq:IdealActiveGasLaw:HigherOrderTheta:app} [or~\meqref{eq:IdealActiveGasLaw:HigherOrderTheta}] for $\vartheta$.

\section{Numerics}
\label{app:Numerics}

\subsection{Overview}

We simulate the Langevin dynamics~\meqref{eq:EOM:x} in
$d=1$ or $2$
using the standard Euler–Maruyama algorithm.
The equations of motion of the self-propulsion processes $\bm v(t)$
depend on the chosen active particle model.
For AOUPs,
$\bm v(t) = \sqrt{2 \Da} \, \bm\eta(t)$ with
\begin{equation}
    \ta \dot{\bm\eta}(t) = -\bm\eta(t) + \bm\zeta(t) \,,
\end{equation}
where $\bm\zeta(t)$ is another Gaussian white-noise process,
independent of $\bm\xi(t)$ in~\meqref{eq:EOM:x}.
For ABPs (only $d=2$),
$\bm v(t) = v_0 \bm n(t)$
with $\bm n(t) = [\cos\theta(t) \; \sin\theta(t)]^\T$
and
\begin{equation}
    \dot{\theta}(t) = \sqrt{\frac{2}{\ta}} \, \zeta(t) \,,
\end{equation}
where $\zeta(t)$ is again an independent Gaussian white-noise process.
For RTPs,
$\bm v(t) = v_0 \bm n_{\kappa(t)}$,
where $\kappa(t)$ is a Poisson counting process with rate $\ta^{-1}$
and $\bm n_0, \bm n_1, \ldots$ is a sequence of independent, uniformly distributed
random variables on the unit sphere in $d$ dimensions.

We choose units such that $\kB T = \gamma = D = 1$.
The
active diffusivity (AOUPs) and self-propulsion speed (RTPs, ABPs) are controlled by the P\'eclet number $\Pe = \sqrt{\Da / D} = v_0 \sqrt{\ta / D d}$.
With this choice, the three active-particle models share the same component-wise self-propulsion velocity correlation
\begin{equation}
    \<v_i(t) \, v_i(s)\> = \frac{\Da}{\tau_a} \e^{-|t-s|/\tau_a}\,.
\end{equation}

The particles are confined in a rectangular box of volume $V = L_1 \cdots L_d$ with reflecting boundary conditions. 
We implement the latter by applying
the sampling method from Ref.~\cite{Behringer2012bds}.

The particle positions are initialized uniformly within the accessible volume.
For AOUPs, the initial $\bm\eta$ is drawn from the marginal steady-state distribution
$\rho_{\mathrm{ss}}(\bm\eta) = (\pi / \ta)^{d/2} \e^{-\ta \bm\eta^2}$.
For ABPs, the initial $\theta$ is drawn uniformly from $[0, 2\pi)$.
For RTPs, $\kappa(0) = 0$.
To reach the steady state,
we then thermalize each system by evolving it for at least $5\ta$.

\subsection{Marginal position distribution}

For each model and configuration, we simulate $N = 10^4$ independent particles with a time step $\Delta t = 10^{-7}$
and take $200$ steady-state snapshots over a period of $\ta$.
The final $50$ snapshots of each trajectory are used to construct the histograms of marginal position distributions in Fig.~\mref{fig:PositionDistribution}.

\subsection{Active ideal gas law}

For the numerical data of the ideal active gas law verification in Fig.~\mref{fig:IdealActiveGasLaw},
we estimate the correlation functions $\< \bm x \cdot \bm v \>$ from the $200$ previously obtained numerical steady-state snapshots of our $10^4$ particles.
We then use the virial representation~\meqref{eq:pressure} of the pressure to calculate $P V$.

\subsection{Numerical work from thermodynamic process}

The thermodynamic work of the expansion processes reported in Fig.~\mref{fig:Process}
involves simulations of AOUPs and RTPs with $\Pe = \sqrt{10}$ and $\Pe = 20$.
The total process time is $\tau = 70\ta$ in the $\Pe = \sqrt{10}$ setups and $\tau = 20\ta$ for $\Pe = 20$.
In the $d=1$ simulations (Fig.~\mref{fig:Process}a and b),
the protocol is
\begin{equation}
    L(t) = L_{\initial} + (L_{\final} - L_{\initial}) \frac{t}{\tau} \,,
\end{equation}
whereas protocols A and B from Eqs.~\meqref{eq:Protocol:LA} and~\meqref{eq:Protocol:LB}, respectively,
were adopted in Fig.~\mref{fig:Process}c.
We choose $L_{\initial} = 0.4$ and $L_{\final} = 0.9$ in all cases
and use a time step of $\Delta t = 5 \times 10^{-9}$ with $N = 10^3$ particles.

\end{document}